\documentclass[twocolumn]{aastex631}
\usepackage{color}
\usepackage[utf8]{inputenc}
\usepackage[T1]{fontenc}
\usepackage{amsmath}

\shortauthors{Zhu, Trussler \& Kewley}

\begin{document}
	
\title{The Nature of Nitrogen Enhanced High Redshift Galaxies}

\author[0000-0002-1333-147X]{Peixin Zhu}
\affiliation{Center for Astrophysics $|$ Harvard \& Smithsonian, 60 Garden Street, Cambridge, MA 02138, USA}

\author[0000-0002-9081-2111]{James Trussler}
\affiliation{Center for Astrophysics $|$ Harvard \& Smithsonian, 60 Garden Street, Cambridge, MA 02138, USA}

\author[0000-0001-8152-3943]{Lisa J. Kewley}
\affiliation{Center for Astrophysics $|$ Harvard \& Smithsonian, 60 Garden Street, Cambridge, MA 02138, USA}
\affiliation{Research School of Astronomy and Astrophysics, Australian National University, Australia}

\email{peixin.zhu@cfa.harvard.edu}

\begin{abstract}

Recent JWST observations have revealed a population of high-redshift galaxies ($z\gtrsim5$) exhibiting unexpectedly bright ultraviolet (UV) nitrogen emission lines. The strong \ion{N}{3}] and \ion{N}{4}] features imply nitrogen-to-oxygen abundance ratios (N/O) as high as $-0.8 \lesssim \log(\mathrm{N/O}) \lesssim 0.4$ in these low-metallicity galaxies ($12+\log(\mathrm{O/H}) \lesssim 8.2$), compared to the local value of $\log(\mathrm{N/O})\approx-1.5$. If confirmed, this level of nitrogen enrichment challenges existing models of nucleosynthesis and galaxy evolution. However, the presence of active galactic nuclei (AGNs) can affect spectral diagnostics, and previous studies often excluded AGN contamination using photoionization models based on local N/O ratios. In this work, we compare nitrogen-enhanced AGN and \ion{H}{2} region models to observed spectra of eight high-redshift galaxies to constrain their nitrogen abundance, excitation source, gas-phase metallicity, ionization parameter, and gas pressure, simultaneously. We find seven galaxies (GHZ9, GS 3073, GN-z9p4, CEERS-1019, GHZ2, GN-z11, and GS-z9-0) are best described by nitrogen-enhanced AGN models, while RXCJ2248-ID is best reproduced by the nitrogen-enhanced \ion{H}{2} model. The presence of AGN does not significantly impact ($\lesssim0.1\,$dex) the derived N/O ratio. We also find that equivalent width (EW)-based diagrams are the most robust UV diagnostic diagrams to distinguish AGNs and star-forming galaxies for situations where the nitrogen abundance is varying. All nitrogen-enhanced galaxies have moderate to high gas pressure ($7.0\leq\log (P/k)\leq9.8$) and high ionization parameter ($\log(U)\gtrsim-2.0$), indicating a dense and compact environment. {\color{black}We suggest that super star clusters containing Wolf-Rayet stars and massive stars} are the most likely contributors to the elevated nitrogen abundance in these galaxies. 
\end{abstract}


\section{Introduction}

The unprecedented spectroscopic capabilities of JWST spectra have uncovered a population of early galaxies ($z\gtrsim5$) that exhibit unexpectedly strong ultraviolet (UV) nitrogen emission lines, including the \ion{N}{3}] quintuplet 1747, 1748, 1749, 1752, and 1754 \AA\, and \ion{N}{4}]$\lambda\lambda1483,86$. Notable examples include GN-z11 \citep{bunker_jades_2023,maiolino_small_2024}, GHZ2 \citep{castellano_jwst_2024}, GHZ9 \citep{napolitano_dual_2024}, and CEERS-1019 \citep{larson_ceers_2023,marques-chaves_extreme_2024}. In combination with UV oxygen lines such as \ion{O}{3}]$\lambda\lambda1661,66$, these nitrogen features imply nitrogen-to-oxygen abundance ratios (N/O) that are elevated by nearly an order of magnitude ($\rm \log (N/O)\gtrsim-0.5$) relative to local galaxies ($\log(\rm N/O)\approx-1.5$), \citet{izotov_heavy-element_1999,2017MNRAS.466.4403N}) at similar low metallicity regimes (12+$\log(\rm O/H)\lesssim7.8$). 

Such elevated N/O ratios in low-metallicity galaxies are surprising, as bright UV nitrogen lines are rarely detected in nearby UV-bright star-forming galaxies \citep{mingozzi_classy_2022,mingozzi_classy_2023}, and they challenge the conventional understanding of nitrogen enrichment. In standard chemical evolution models, nitrogen enrichment {\color{black}consists of a primary nucleosynthesis (metallicity-independent) and a secondary nucleosynthesis (metallicity-dependent). Sources producing nitrogen include core-collapse supernovae \citep{van_zee_abundances_1998} and intermediate-mass stars via hot-bottom burning and dredge-up \citep{renzini_advanced_1981}.} Because massive stars evolve more quickly than intermediate-mass stars, primary nucleosynthesis is expected to dominate at low metallicities, resulting in a relatively flat and low N/O plateau at 12 + $\log(\rm O/H) \lesssim 7.8$, while the secondary nucleosynthesis begins to dominate over primary nucleosynthesis at 12+$\log(\rm O/H)\gtrsim8.0$ \citep{israelian_galactic_2004,karakas_dawes_2014,2017MNRAS.466.4403N,cameron_nitrogen_2023}.

Several studies have investigated the possible formation scenario of the nitrogen excess phenomenon. Proposed sources for nitrogen enrichment include Wolf-Rayet (WR) stars \citep{watanabe_empress_2024}, supermassive stars ($M>1000M_{\odot}$) \citep{charbonnel_n-enhancement_2023, senchyna_gn-z11_2023,marques-chaves_extreme_2024}, super star clusters \citep{pascale_nitrogen-enriched_2023}, pollution from Pop III stars \citep{maiolino_jades_2024}, AGB stars, and tidal disruption of stars from encountering black holes \citep{cameron_nitrogen_2023,johnson_empirical_2023,watanabe_empress_2024}. In addition to the peculiar nucleosynthesis, the enhanced N/O may could also result from pristine gas accretion that reduces the metallicity while keeping the N/O ratio fixed {\color{black}\citep{stiavelli_what_2025}, or preferential ejection of oxygen \citep {rizzuti_high_2025}.} 
While all these mechanisms could be tailored to predict the elevated N/O ratio in the nitrogen-enhanced galaxies, it is hard to constrain the actual dominant mechanism with current observations.

While the mechanism responsible for nitrogen excess remains uncertain, the presence of active galactic nuclei (AGNs) and shocks may also affect the N/O ratio estimation. For instance, \citet{maiolino_small_2024} found that the \ion{N}{4}]$\lambda\lambda1483,86$ doublet ratio {\color{black} and the \ion{N}{3}] quintuplet in GN-z11 indicate electron densities of $n_{\rm e}\gtrsim10^6\rm \,cm^{-3}$ or higher, consistent with the physical conditions expected in the AGN broad-line region. At such high densities, they further suggested that only a small amount of nitrogen mass is required to reproduce the observed line strengths, provided that the nitrogen is concentrated in the broad-line regions (BLRs) rather than distributed throughout the galaxy.} In addition, \citet{flury_new_2024} found that shock models provide a better match to the observed UV emission line ratios in N-enhanced galaxies than \ion{H}{2} or AGN photoionization models assuming the same local nitrogen abundance scaling relation, suggesting that shocks can reduce the derived nitrogen abundance into better agreement with local values. 

However, identifying AGN activity at high redshift is challenging, particularly in the absence of broad hydrogen recombination lines that trace AGN broad-line regions or broad oxygen emission lines indicative of AGN-driven outflows {\color{black}\citep[e.g.,][]{mullaney_narrow-line_2013,harrison_kiloparsec-scale_2014,veilleux_cool_2020}.} In the lack of broad emission line features and {\color{black}very high-ionization lines, such as [Ne~V]$\lambda3426$ that requires photons with energies $\gtrsim97\,$eV}, the most widely used method to assess AGN contamination is through comparing the observed emission line ratios with predictions from the AGN and \ion{H}{2} photoionization models on the UV diagnostic diagrams \citep[e.g.][]{bunker_jades_2023,senchyna_gn-z11_2023,marques-chaves_extreme_2024,napolitano_dual_2024,castellano_jwst_2024}. 

UV diagnostic diagrams have been studied for decades to facilitate the classification of star-forming and Seyferts galaixes {\color{black}at low and high redshift} \citep{villar-martin_ionization_1997,allen_ultraviolet_1998,groves_dusty_2004,nagao_gas_2006,feltre_nuclear_2016,nakajima_vimos_2018,dors_semi-empirical_2019,hirschmann_synthetic_2019,nakajima_diagnostics_2022,mingozzi_classy_2023}. However, these UV diagnostic diagrams are developed using photoionization models that assume a local nitrogen abundance, which is not consistent with the enhanced N/O ratios reported in these N-excess galaxies. Therefore, the separation criteria proposed in previous studies may not be applicable to nitrogen-enhanced galaxies.

In this paper, we overcome the inconsistency in nitrogen abundance between models and observations by developing nitrogen-enhanced photoionization models and comparing them to the observed UV emission line properties of the N-excess galaxies at high redshift. This approach allows us to simultaneously constrain the nitrogen abundance, dominant excitation mechanisms, gas-phase metallicity, and ionization parameter in these early galaxies. We further explore how the presence of AGNs influences nitrogen abundance estimation and discuss the implications for the origin of nitrogen enrichment.

This paper is organized as follows: Section 2 describes the observational data, and Section 3 introduces the nitrogen-enhanced models. In Section 4, we compare the nitrogen-enhanced models with observational data on UV diagnostic diagrams. Section 5 presents the best-fit models tailored for each high-redshift nitrogen-enhanced galaxy. In Section 6, we discuss the dominant excitation source and the nitrogen enrichment mechanisms for nitrogen-enhanced galaxies. Section 7 summarizes the conclusions. {\color{black}We follow the convention to adopt vacuum wavelengths for space-based JWST observation in both UV and optical wavelengths through out this work.}

\section{Observational Data}\label{sec:data}

{\color{black}To investigate the nitrogen enhancement phenomenon in the early universe, we assemble a “N-enhanced sample”—a collection of galaxies at $z\gtrsim5$ that exhibit significantly elevated nitrogen-to-oxygen abundance ratios ($\log(\rm N/O)\gtrsim-1.0$) compared to the local universe {\color{black} at low metallicity regime (12+$\log(\rm O/H)\lesssim8.2$)}, and have $>3\sigma$ UV \ion{N}{4}] emission lines detection.} This sample includes GN-z11 \citep{bunker_jades_2023,maiolino_small_2024}, GHZ2 \citep{castellano_jwst_2024}, GHZ9 \citep{napolitano_dual_2024}, CEERS-1019 \citep{larson_ceers_2023,marques-chaves_extreme_2024}, GN-z9p4 \citep{schaerer_discovery_2024}, GS 3073 \citep{ji_ga-nifs_2024}, GS-z9-0 \citep{curti_jades_2024}, and RXCJ2248-ID \citep{topping_metal-poor_2024}. 

Table~\ref{tab:1} presents the redshift, N/O ratio, C/O ratio, gas-phase metallicity, electron density (when available), {\color{black}electron temperature,} the $A_V$, and the emission line fluxes and UV equivalent widths measurements of each galaxy in the ``N-enhanced sample'' collected from the corresponding reference. To facilitate the best-fit model comparison in Section~\ref{sec:5}, we also include optical emission line fluxes for [O~II]$\lambda\lambda$3727,30, [Ne~III]$\lambda$3870, H$\delta$, H$\gamma$,  [\ion{O}{3}]$\lambda$4364, \ion{He}{2}$\lambda$4687, H$\beta$,  [\ion{O}{3}]$\lambda$5008, H$\alpha$, [N~II]$\lambda$6585, and  [\ion{S}{2}]$\lambda\lambda$6718,33 from the corresponding reference. For galaxies observed with both PRISM spectra and NIRSpec gratings, we adopt measurements from the PRISM data, {\color{black}which provide broader wavelength coverage and more reliable continuum estimates for EW measurements. This choice does not affect our conclusion, as emission line ratios from PRISM and the NIRSpec gratings generally agree within $\lesssim0.2\,$dex for lines detected at the $>3\sigma$ level.}

The emission line fluxes listed in Table~\ref{tab:1} are compiled from the literature and are presented without correction for dust extinction, as most of these galaxies appear to be nearly dust-free. {\color{black}To illustrate the effect of a worst-case (upper-limit) dust extinction on the ``N-enhanced sample'' in the comparison with the theoretical unreddened line ratios in Section~\ref{sec:4}, we adopt the extinction curve from \citet{cardelli_relationship_1989} and plot the vector for $A_V$=5.0 mag with $R_V=3.1$ using a black arrow on each UV diagram.} 

\setlength{\tabcolsep}{0.6pt}
\begin{deluxetable*}{c|c|c|c|c|c|c|c|c|c|c}[hbt]
\small
\centering
\tablewidth{0pc}
\tablecaption{Properties of the ``N-enhanced sample'' \label{tab:1}}
\tablenum{1}
\tablehead{
\colhead{Galaxy} &
\colhead{GN-z11} &
\colhead{GN-z11} &
\colhead{GHZ2} &
\colhead{GHZ9} &
\colhead{CEERS-1019} &
\colhead{CEERS-1019} &
\colhead{GN-z9p4} &
\colhead{GS 3073} &
\colhead{GS-z9-0} &
\colhead{RXCJ2248-ID}
}
\startdata
$z$  & 10.60 & 10.60 & 12.34 & 10.15 & 8.68 & 8.68 & 9.38 & 5.55 & 9.43 & 6.11 \\
$\log$(N/O) & $\gtrsim-0.4$ & $\gtrsim-0.4$ & $-0.29$ & $-0.08$ & $--$ & -0.13$\pm$0.11 & -0.59$\pm$0.24 & 0.42$^{+0.13}_{-0.10}$ & -0.93$\pm$0.28 & $-0.39^{+0.11}_{-0.10}$ \\
$\log$(C/O)  & $--$ & $--$ & $-0.94$ & $-0.96$ & $--$ & -0.75$\pm$0.11 & $<-1.18$ & $-0.38^{+0.13}_{-0.11}$ & $-0.9^{+0.27}_{-0.19}$ & $-0.83$ \\
12+$\log(\rm O/H)$  & 7.59-7.76 & $--$ & 7.26 & 6.76-7.76 & 7.66$\pm$0.51 & 7.70$\pm$0.18 & 7.37$\pm$0.15 & 8.00$^{+0.12}_{-0.09}$ & 7.49$^{+0.11}_{-0.15}$ & 7.43$^{+0.17}_{-0.09}$ \\
$\log (n_{\rm e}/\rm cm^3)$  & $--$ & $\gtrsim6.0$ & 5.0\tablenotemark{\scriptsize a} & 5.0\tablenotemark{\scriptsize a} & $\geq$4 & 4-5 & 2.0\tablenotemark{\scriptsize a} & 5.7 & 2.8 & 4.8-5.5 \\
$T_e$ ($10^4$ \rm K)  & $--$ & $--$ & 3.0\tablenotemark{\scriptsize a} & 1.5\tablenotemark{\scriptsize a} & 1.86 & 1.8$\pm$0.1 & 2.3$\pm$0.4\tablenotemark{\scriptsize b} & 1.7-2.1 & 2.4$\pm$0.2 & 2.46$\pm$0.26 \\
Av & 0.17$\pm$0.03 & $--$ & $0.04^{+0.07}_{-0.03}$ & $--$\tablenotemark{\scriptsize c} & 0.4$\pm$0.2 & 0.53$\pm$0.06 & 0.3$\pm$0.2 & 0.49$\pm$0.01 & 0.00$\pm$0.07 & $0.02^{+0.03}_{-0.02}$ \\
References\tablenotemark{\dag} & (1) & (2) & (3) & (4) & (5) & (6) & (7) & (8) & (9) & (10) \\
\hline
Flux Unit (erg\,s$^{-1}$\,cm$^{-2}$) & $10^{-19}$ & $10^{-19}$ & $10^{-19}$ & $10^{-19}$ & $10^{-18}$ & $10^{-18}$ & $10^{-18}$ & $10^{-18}$ & $10^{-19}$ & $10^{-19}$ \\
\hline
\ion{N}{5}$\lambda\lambda1239,43$ &  & <3.5 &  &  & <1.74 & <1.44 &  &  &  &  \\
\ion{N}{4}]$\lambda\lambda$1483,86 & $8.6\pm1.8$ & $12.0\pm1.7$ & $6.9\pm0.6$ & $12.5\pm2.0$ & $3.36\pm0.44$ & $3.75\pm0.4$ & $2.913\pm0.53$ & $54.1\pm0.9$ & $1.34\pm0.64$ & $29.4\pm1.0$ \\
\ion{C}{4}$\lambda\lambda$1548,51 &  & $3.2\pm0.8$ & $25.7\pm0.6$ & $17.3\pm1.9$ &  & $2.1\pm0.42$ & <1.56 & $9.4\pm1.1$ & $4.2\pm0.66$ & $61.7\pm0.9$ \\
\ion{He}{2}$\lambda$1640 &  & $6.0\pm1.4$ & $2.7\pm1.6$ & $4.5\pm2.0$ &  & <1.2 &  & $13.9\pm0.9$ & $1.72\pm0.24$ & $5.9\pm0.4$ \\
\ion{O}{3}]$\lambda\lambda$1661,66 &  &  & $7.2\pm1.5$ & $6.9\pm1.9$ &  & $1.64\pm0.32$ & <1.08 & $3.8\pm0.9$ & $3.33\pm0.46$ & $23.1\pm0.5$ \\
\ion{N}{3}]$\lambda$1747$\_$54 & $7.3\pm1.0$ & $10.3\pm1.7$ & $3.4\pm0.9$ & $7.9\pm1.2$ &  & $0.73\pm0.3$ & <1.32 & $20.7\pm0.8$ & $1.16\pm0.39$ & $4.5\pm0.4$ \\
\ion{C}{3}]$\lambda\lambda$1907,09 & $10.4\pm1.6$ & $9.9\pm0.9$ & $9.1\pm0.2$ & $11.0\pm1.2$ & <1.7 & $2.43\pm0.36$ & <1.11 & $21.0\pm0.4$ & $4.33\pm0.41$ & $22.2\pm0.7$ \\
{[}\ion{O}{2}]$\lambda\lambda$3727,30 & $9.1\pm0.4$ & $7.72\pm1.3$ & $2.7\pm1.1$ & $1.99\pm0.39$ & $2.94\pm0.25$ &  & $0.683\pm0.2$ & $7.0\pm0.2$ & $0.89\pm0.17$ & $1.1\pm0.4$ \\
{[}\ion{Ne}{3}]$\lambda$3870 & $7.9\pm0.7$ & $10.0\pm0.7$ & $6.4\pm0.8$ & $4.11\pm0.41$ & $3.5\pm0.27$ &  &  & $20.4\pm0.2$ & $2.52\pm0.24$ & $19.8\pm0.4$ \\
H$\delta$ & $5.7\pm0.5$ & $5.1\pm0.8$ &  & $1.41\pm0.27$ & $1.14\pm0.37$ &  & $0.976\pm0.223$ & $9.32\pm0.14$ & $1.88\pm0.16$ & $7.2\pm0.3$ \\
H$\gamma$ & $10.7\pm1.1$ & $9.2\pm2.8$ &  & $3.79\pm0.49$ & $2.6\pm0.39$ &  & $1.76\pm0.18$ & $13.1\pm0.2$ & $3.18\pm0.17$ & $13.0\pm0.2$ \\
{[}\ion{O}{3}]$\lambda$4364 & $2.3\pm0.7$ &  &  & $2.9\pm0.5$ & $1.16\pm0.38$ &  & $0.691\pm0.16$ & $5.3\pm0.2$ & $1.09\pm0.16$ & $10.7\pm0.2$ \\
\ion{He}{2}$\lambda$4687 &  &  &  &  &  &  &  & $7.6\pm0.3$ &  & <0.9 \\
H$\beta$ &  &  &  &  & $7.32\pm0.72$ &  & $3.314\pm0.14$ & $37.9\pm0.2$ & $6.76\pm0.19$ & $25.4\pm0.2$ \\
{[}\ion{O}{3}]$\lambda$5008 &  &  &  &  & $39.58\pm0.84$ &  & $16.25\pm0.18$ & $146.0\pm0.2$ & $29.22\pm0.25$ & $166.9\pm0.2$ \\
H$\alpha$ &  &  &  &  &  &  &  & $123.7\pm0.3$ &  & $64.7\pm0.2$ \\
{[}\ion{N}{2}]$\lambda$6585 &  &  &  &  &  &  &  &  &  & $1.6\pm0.2$ \\
{[}\ion{S}{2}]$\lambda\lambda$6718,33 &  &  &  &  &  &  &  & $1.26\pm0.14$ &  & <1.4 \\
\hline
\multicolumn{11}{c}{Rest-frame equivalent widths measurements for UV emission lines (\AA)}\\
\hline
EW(\ion{N}{5}) &  & <19.0 &  &  & $3.9\pm2.3$ &  &  &  &  &  \\
EW(\ion{N}{4}]) & $9.0\pm1.1$ &  & $12.1\pm1.2$ & $47.0\pm8.0$ & $7.96\pm1.47$ &  & $31.8\pm6.46$ &  & $2.0\pm1.0$ & $17.35\pm0.7$ \\
EW(\ion{C}{4}) &  &  & $45.8\pm1.2$ & $65.0\pm7.0$ &  &  & <19.1 &  & $7.0\pm1.0$ & $34.1\pm0.8$ \\
EW(\ion{He}{2}) &  &  & $4.9\pm3.1$ & $18.0\pm8.0$ &  &  &  &  & $7.0\pm1.0$ & $5.6\pm0.4$ \\
EW(\ion{O}{3}]) &  &  & $13.7\pm4.8$ & $28.0\pm8.0$ &  &  & <14.7 &  & $9.0\pm1.0$ & $18.8\pm0.5$ \\
EW(\ion{N}{3}]) & $9.4\pm1.0$ &  & $6.8\pm2.1$ & $33.0\pm5.0$ &  &  & <19.7 &  & $3.0\pm1.0$ & $4.3\pm0.5$ \\
EW(\ion{C}{3}]) & $14.3\pm2.3$ &  & $25.6\pm12.5$ & $48.0\pm5.0$ & <6.04 &  & <18.9 &  & $14.0\pm2.0$ & $21.7\pm0.6$ \\
\enddata
\tablenotetext{\dag}{(1)\citet{bunker_jades_2023}; (2)\citet{maiolino_small_2024}; (3)\citet{castellano_jwst_2024}; (4)\citet{napolitano_dual_2024}; (5)\citet{larson_ceers_2023}; (6)\citet{marques-chaves_extreme_2024}; (7)\citet{schaerer_discovery_2024}; (8)\citet{ji_ga-nifs_2024}; (9)\citet{curti_jades_2024}; (10)\citet{topping_metal-poor_2024}}
\tablenotetext{a}{No direct measurement is available. The listed value is the assumed value adopted in abundance and metallicity estimates.}
\tablenotetext{b}{Estimated assuming $n_{\rm e}=100\,\rm cm^{-3}$. At $n_{\rm e}\gtrsim10^5\,\rm cm^{-3}$, the inferred $T_e$ would be lower and lead to a higher O/H and even more higher N/O and C/O ratios \citep{schaerer_discovery_2024}.}
\tablenotetext{c}{The value derived from the UV slope under the assumption of no AGN contribution is 4.96$\pm3.1$. However, X-ray observation suggests that this galaxy does host an AGN \citep{kovacs_candidate_2024,napolitano_dual_2024}.}
\end{deluxetable*}
\vspace{-2em} 

To evaluate our \ion{H}{2} photoionization models, we incorporate observational data from the Cosmic Origins Spectrograph (COS) Legacy Spectroscopic Survey {\color{black}(CLASSY, \citet{berg_cos_2022,james_classy_2022}),} which comprises a sample of 45 UV-bright, star-forming galaxies in the local universe ($0.002<z<0.182$). The CLASSY galaxies span a broad range of stellar masses ($6.98<\log \rm(M_{\star}/M_{\odot})<10.06$), star formation rates ($-2<\log\rm(SFR(M_{\odot}/yr)<2$), and gas metallicity ($7.0<12+\log\rm(O/H)<8.8$), and are all confirmed to be star-forming galaxies based on their positions on the standard optical diagnostic diagrams \citep{baldwin_classification_1981,veilleux_spectral_1987-1,kewley_host_2006}. Although most galaxies in this sample do not exhibit strong \ion{N}{3}] multiplet detection (except for one galaxy, Mrk 996) \citep{mingozzi_classy_2022}, these galaxies can serve as a benchmark for the location of star-forming galaxies in the UV diagnostic diagrams. We adopt the dereddened UV emission line flux measurements from \citet{mingozzi_classy_2022}. 

We also include a sample of $z<4$ AGNs with available UV emission line observations from \citet{dors_semi-empirical_2019} to evaluate the AGN photoionization models. We adopt the emission-line fluxes of {\color{black}N~V$\lambda\lambda$1239,43, \ion{He}{2}$\lambda$1640, \ion{C}{3}]$\lambda\lambda$1907,09, \ion{C}{4}$\lambda\lambda$1548,51} for 77 Type 2 AGNs from Table 1 in \citet{dors_semi-empirical_2019}. These 77 Type 2 AGNs consist of 9 local Type 2 Seyferts ($z<0.03$), 7 X-ray selected Type 2 quasars ($1.5<z<3.6$), and 61 radio galaxies at $1.2<z<4.0$. These UV observations are compiled from \citet{kraemer_iue_1994,de_breuck_sample_2000,nagao_gas_2006,bornancini_imaging_2007,matsuoka_chemical_2009} and \citet{matsuoka_mass-metallicity_2018}. The UV fluxes of this sample have not been corrected for dust extinction {\color{black}due to the lack of dust measurements. The $A_V$=5.0 mag arrow on each UV diagnostic diagram marks a worst-case dust extinction scenario.}

\section{theoretical models}\label{sec:model}

\subsection{Stellar Ionizing Spectra}\label{sec:3.1}

{\color{black}We adopt single burst stellar ionizing spectra generated by Flexible Stellar Population Synthesis (FSPS, \citet{conroy_propagation_2009,conroy_propagation_2010}) using the MESA Isochrones and Stellar Tracks (MIST, \citet{paxton_modules_2011,paxton_modules_2013,paxton_modules_2015,dotter_mesa_2016,choi_mesa_2016}) with stellar rotation rate of $v/v_{\rm crit}=0.4$ and the Chabrier initial mass function \citep{chabrier_galactic_2003} with a mass range of $10\leq M_{\star}/M_{\odot}\leq300$. Adopted stellar spectral library includes the MILES \citep{sanchez-blazquez_medium-resolution_2006,falcon-barroso_updated_2011}, CMFGEN Wolf-Rayet (WR) Stars spectra \citep{smith_realistic_2002}, and WM-Basic hot star spectra \citep{pauldrach_wm-basic_2012}. The WR stars are identified when stars have $T_{\rm eff}>10^4\,$K and the surface hydrogen mass fraction $X_{\rm surf}<0.3$. The WN and WC subtypes are classified based on the surface C/N number ratio. WR stars with C/N$>1$ are labeled as WC and vice versa. The stellar ionizing spectra are taken from snapshots at 3\,Myr, when these single-burst models have the hardest stellar ionizing spectra and have comparable hardness with those of binary-star models from BPASS \citep{choi_evolution_2017}.

We use the same stellar ionizing spectra for both the `N-normal' and `N-enhanced' \ion{H}{2} region models, without adjusting the stellar N/O ratio separately for the `N-enhanced' \ion{H}{2} models.} This assumption does not impact our conclusions, as \citet{grasha_stromlo_2021} demonstrated that variations in elemental abundance ratios have a relatively minor effect on stellar evolutionary tracks compared to rotation and metallicity.

\subsection{AGN Ionizing Spectra}

{\color{black}A detailed description of the AGN ionizing spectra is provided in \citet{zhu_new_2023}, which we briefly summarize here. We adopt the AGN ionizing spectra generated from OXAF \citep{thomas_physically_2016}, a simplified version of the AGN radiation model OPTXAGNF \citep{done_intrinsic_2012,jin_combined_2012} that models the continuum emission radiated from the AGN accretion disk and a Comptonizing corona surrounding a central rotating black hole. These ionizing spectra models are valid for AGNs in thin-disk accretion modes, typically in the range of $10^6\lesssim M_{\rm BH}/M_{\odot}\lesssim10^9$ and $0.05\lesssim L/L_{\rm Edd} \lesssim0.3$ \citep {novikov_astrophysics_1973,done_intrinsic_2012,thomas_physically_2016}. These black hole mass and accretion rate ranges are consistent with those inferred from JWST-discovered high-$z$ AGNs \citep{harikane_jwstnirspec_2023,matthee_little_2024,juodzbalis_jades_2025}.} The most important parameter in the OXAF model is $E_{\rm peak}$, the peak energy of the accretion disk thermal emission \citep{thomas_physically_2016,zhu_new_2023}, which reflects the hardness of the AGN ionizing spectra. We adopt $E_{\rm peak}$ in the range of $-2.0\leq\log (E_{\text{peak}}/\text{keV})\leq-1.0$ with a 0.25 dex interval in the AGN photoionization models.

\subsection{Photoionization Models}
We use MAPPINGS version 5.2 \citep{binette_radiative_1985,sutherland_cooling_1993,dopita_new_2013,sutherland_mappings_2018} to model the \ion{H}{2} region and AGN narrow-line region. MAPPINGS version 5.2 utilizes atomic data from the CHIANTI version 10 database \citep{del_zanna_chiantiatomic_2021} for the 30 lightest elements. {\color{black}We consider isobaric (constant-pressure) structure in all models, which naturally permits density and temperature variations as observed in the high-$z$ galaxies.} Dust depletion is considered in all theoretical models, using the dust depletion factors from \citet{jenkins_unified_2009} where the depletion value for Fe is $-1.5\,$dex. This value is chosen based on the Wide Integral Field Spectrograph observations of local and Magellanic Cloud \ion{H}{2} regions (Nicholls et al., in preparation). {\color{black}For metal-poor (12+$\log(\rm O/H)\leq8.4$) galaxies with little or no dust ($A_V<0.05$), adopting dusty models does not affect our results, because dusty and dust-free models occupy nearly identical loci on the UV diagnostic diagrams (Section~\ref{sec:4}) and optical diagnostic diagram \citep{zhu_new_2023} at 12+$\log(\rm O/H)\leq8.4$.} 

For the \ion{H}{2} and AGN models with normal (local) nitrogen abundance (called `N-normal' models in the following sections), we adopt the abundance scaling relations from \citet{2017MNRAS.466.4403N}, which use the local B-stars abundance set from \citet{nieva_present-day_2012} and fit the abundance scaling relation based on the observations of nearby \ion{H}{2} regions \citep[e.g.][]{izotov_heavy-element_1999,israelian_galactic_2004,spite_first_2005,fabbian_co_2009,nieva_present-day_2012}. These observations suggest an average N/O ratio of $\log(\rm N/O)\approx-1.5$ at 12+$\log(\rm O/H)\lesssim8.0$, with a scatter ranging from approximately $-2.3\lesssim\log(\rm N/O)\lesssim-1.0$ \citep{izotov_heavy-element_1999, israelian_galactic_2004,spite_first_2005}. 

To compare with the high-redshift nitrogen-enhanced galaxies, we develop a set of \ion{H}{2} and AGN models with their nitrogen abundance tailored to match the reported elevated N/O values. These models are built by adopting the same setting as `N-normal' models, but simply changing the N/O ratio to vary with 0.1 dex intervals in the range of $-0.8\lesssim\log(\rm N/O)\lesssim0.4$. These models are referred to as `N-enhanced' models in the following sections.

In all \ion{H}{2} and AGN photoionization models, we vary the gas-phase metallicity in the range of $7.2\leq$12+$\log(\rm O/H)\leq8.7$, ionization parameter in the range of $-4.0\leq\log(U)\leq-0.5$, {\color{black}and gas pressure in the range of 5.0$\leq\log{(P/k)}\leq$9.8 (corresponding to gas density of $1.0\lesssim \log(n/\rm cm^{-3})\lesssim6.0$ at $T\approx(1-2.5)\times10^4$\,K). The high-pressure limit is chosen to encompass the extremely high $n_{\rm e}$ reported in some galaxies of our ``N-enhanced sample'' \citep[e.g.][]{maiolino_small_2024,topping_metal-poor_2024}.} 

\section{Comparison of Nitrogen Enhanced Photoionization Models with UV Observations}\label{sec:4}

In this section, we compare \ion{H}{2} and AGN photoionization models with UV observations of both local and high-redshift galaxies on the UV diagnostic diagrams. Our objectives are twofold. First, to assess the efficacy of \ion{H}{2} and AGN models in reproducing the observed spectra of star-forming galaxies and Seyfert galaxies, respectively. Second, to evaluate the diagnostic power of these UV diagrams in constraining the dominant excitation sources, gas metallicity, ionization parameter, gas pressure, and nitrogen abundance in the observed galaxies.
 
\begin{figure*}[htb]
\epsscale{1.}
\plotone{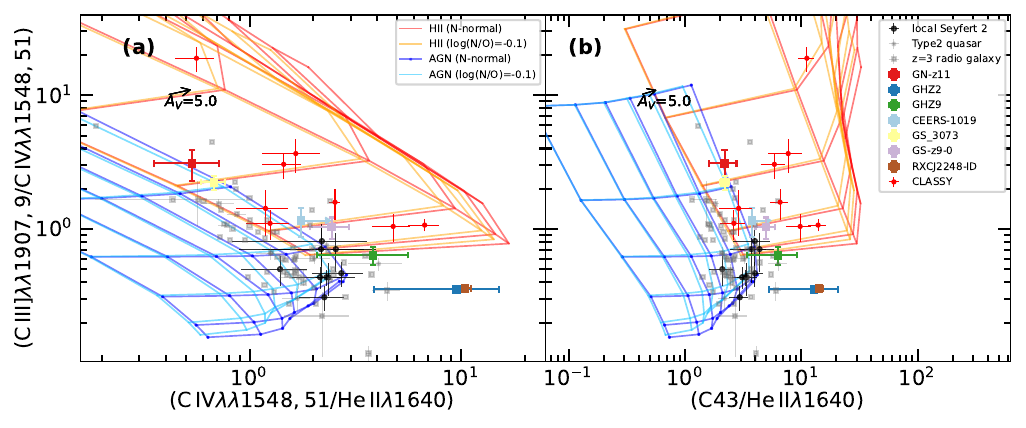}
\caption{Comparison of \ion{H}{2} and AGN photoionization models with observations on UV diagnostic diagrams. \ion{H}{2} models at $\log \rm(P/k)=7.8$ with `N-normal' and `N-enhanced' nitrogen abundance are presented in red and orange grids. AGN models at $\log \rm(P/k)=8.4$ and $\log(E_{\text{peak}}/\text{keV})= -1.0$ are presented in blue (`N-normal') and deepskyblue (`N-enhanced') grids. Each model grid consists of constant metallicity lines (from left to right: $12+\log(\rm O/H)=7.30, 7.70, 8.00, 8.28, 8.43, 8.70$ for AGN models; $12+\log(\rm O/H)=7.15, 7.75, 8.15, 8.43, 8.63$ for \ion{H}{2} models) and constant ionization parameter lines (from top to bottom: $\log(\rm U)=-3.0,-2.5,-2.0,-1.5,-1.0,-0.5$). Observations for local Seyfert 2 galaxies, Type 2 quasars, and $z<4$ radio galaxies are presented with solid black circles, solid grey dots, and solid grey squares. CLASSY galaxies are shown as red circles, and $z>5$ nitrogen-enhanced galaxies are highlighted as large colored squares, with color indicating individual sources. Black arrows in each diagram indicate {\color{black}a worst-case dust impact ($A_V=5.0\,$ mag at $R_V=3.1$)} for the UV observations that are not corrected for dust extinction ($z<4$ AGNs and $z>5$ nitrogen-enhanced galaxies). {\color{black}These diagrams show that our \ion{H}{2} models can reproduce the CLASSY star-forming galaxies and AGN models can reproduce the $z<4$ AGNs, which provide a validated baseline for interpreting high-redshift N-enhanced galaxies.}
 \label{fig:1}}
\end{figure*}

We begin by examining the \ion{C}{4}/\ion{He}{2}$-$\ion{C}{3}]/\ion{C}{4} and the C43/\ion{He}{2}-\ion{C}{3}]/\ion{C}{4} diagrams (C43=\ion{C}{3}]+\ion{C}{4}), introduced by \citet{villar-martin_ionization_1997,allen_ultraviolet_1998, nagao_gas_2006,feltre_nuclear_2016,nakajima_vimos_2018}. {\color{black}Although \ion{C}{4} and \ion{He}{2} can be affected by stellar-wind emission, ISM absorption, and resonant scattering, these diagrams are widely used to distinguish \ion{H}{2} regions from AGN in rest-UV observations. We compare our \ion{H}{2} and AGN photoionization models with all UV observations on these diagrams in Figure~\ref{fig:1}.} We find that our `N-normal' \ion{H}{2} model can account for most of the CLASSY galaxies. This aligns with optical analysis that all CLASSY galaxies are classified as star-forming galaxies on the standard optical diagnostic diagrams \citep{mingozzi_classy_2023}. The `N-normal' AGN model adopting a relatively hard radiation field $\log (E_{\text{peak}}/\text{keV})= -1.0$ can predict most UV observations for AGNs at $z<4$. For both mechanisms, the `N-enhanced' model exhibits a distribution very similar to the `N-normal' model in Figure~\ref{fig:1}, {\color{black} confirming} nitrogen abundance has little impact on diagnostics that do not involve nitrogen emission lines.

On the UV diagnostic diagrams shown in Figure~\ref{fig:1}, most high-redshift Nitrogen enhanced galaxies are located at the boundary between \ion{H}{2} and AGN models, which indicates these galaxies are either extremely metal-poor (12+$\log(\rm O/H)\approx7.2$) star-forming galaxies, or AGNs that exhibit relatively soft radiation fields ($\log (E_{\text{peak}}/\text{keV})\leq-1.7$) with sub-solar metallicity, or a mixing of star formation and AGN with moderate metal-poor gas ($7.2\lesssim$12+$\log(\rm O/H)\lesssim7.9$). Additional observational information is needed to constrain the nature of their dominant excitation source. 

\begin{figure}[htb]
\epsscale{1.1}
\plotone{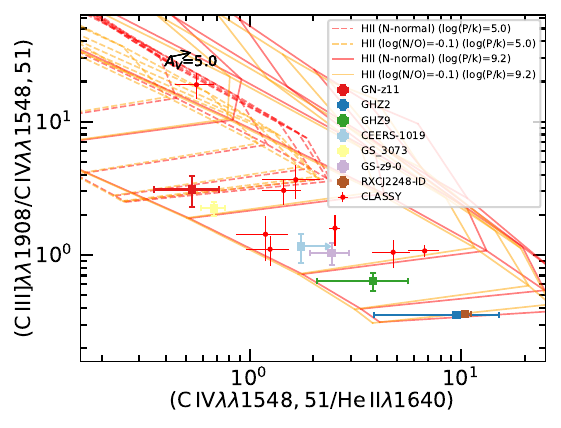}
\caption{Comparison of \ion{H}{2} photoionization models at $\log \rm(P/k)=5.0$ (dashed grids) and $\log \rm(P/k)=9.2$ (solid grids). Model grid parameter ranges and symbols are the same as in Figure~\ref{fig:1}. {\color{black}Only high-pressure ($5.0<\log \rm(P/k)\lesssim9.2$) \ion{H}{2} models can reproduce CLASSY galaxies and high-$z$ `N-enhanced' galaxies.} \label{fig:2}}
\end{figure}

\begin{figure*}[htb]
\epsscale{1.2}
\plotone{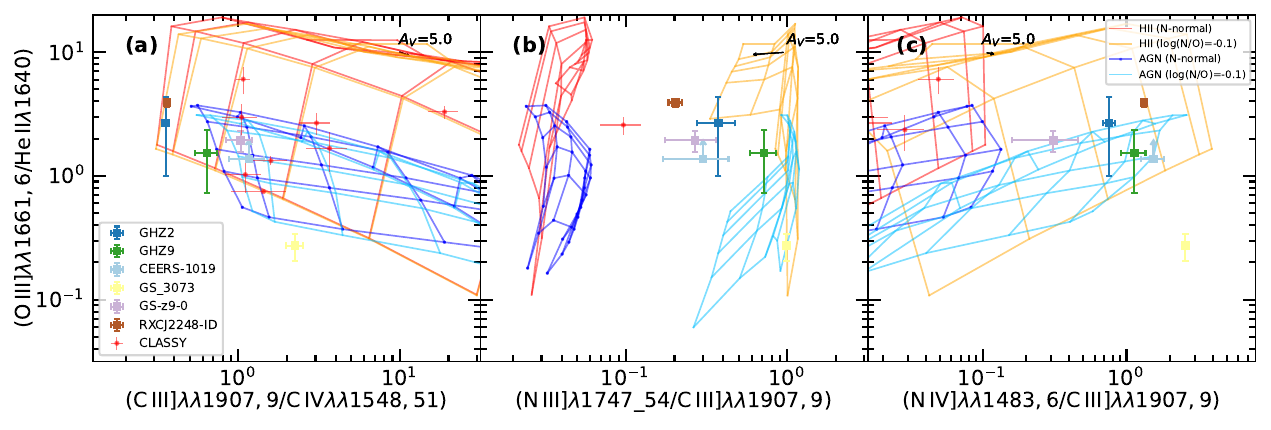}
\caption{Comparison of \ion{H}{2} and AGN photoionization models with observations on OIII/HeII-based UV diagnostic diagrams. Model grid parameter ranges and symbols are the same as in Figure~\ref{fig:1}, except that HII models now have gas pressures of $\log \rm(P/k)=9.2$ and AGN models now have a softer radiation field $\log (E_{\text{peak}}/\text{keV})=-2.0$. These model parameters are adjusted to better characterize the high-$z$ nitrogen-enhanced galaxies. {\color{black} Only models with elevated N/O ratio can reproduce the bright \ion{N}{3}] and \ion{N}{4}] emission lines observed in high-$z$ `N-enhanced' galaxies.}
 \label{fig:3}}
\end{figure*}

Remarkably, the low redshift, UV-bright star-forming galaxies observed in the CLASSY survey exhibit significantly higher gas pressure compared to typical values found in optical-selected star-forming galaxies ($\log \rm(P/k)\approx4.0-5.0$, see \citet{kewley_understanding_2019}).  {\color{black}This is shown by Figure~\ref{fig:2}, where both \ion{H}{2} models with gas pressure of $\log \rm(P/k)=5.0$ and $\log \rm(P/k)=9.2$ are shown on the \ion{C}{4}/\ion{He}{2}$-$\ion{C}{3}]/\ion{C}{4} diagram.} The \ion{H}{2} model with $\log \rm(P/k)=5.0$ fails to reproduce the observed line ratios of CLASSY galaxies on the diagram, despite its success in modeling star-forming galaxies on the standard optical diagnostic diagram. To reproduce the CLASSY galaxies on the \ion{C}{4}/\ion{He}{2}$-$\ion{C}{3}]/\ion{C}{4} diagram, {\color{black}\ion{H}{2} models need to have gas pressure in the range $5.4\lesssim\log \rm(P/k)\lesssim7.8$, corresponding to electron densities of approximately $10^2\lesssim n_{\rm e}(\rm cm^{-3})\lesssim10^4$ at $T_e\approx(1-1.5)\times10^4\,$K. Our electron density estimations for CLASSY galaxies are consistent with the values estimated from optical [\ion{S}{2}] in \citet{mingozzi_classy_2023,arellano-cordova_classy_2025}. Figure~\ref{fig:2} also suggests that reproducing the high-$z$ `N-enhanced' galaxies with \ion{H}{2} models would require extremely high gas pressure ($\log \rm(P/k)\gtrsim 7.0$) and very low metallicity ($12+\log(\rm O/H)\lesssim 7.7$).}

{\color{black}We then compare the models with observations on three \ion{O}{3}]/\ion{He}{2} based UV diagrams in Figure~\ref{fig:3}. We find that AGN models with $\log (E_{\text{peak}}/\text{keV})= -1.0$ deviate from the high-$z$ `N-enhanced' galaxies on these diagrams, and softer AGN ionizing spectra ($\log(E_{\rm peak}/{\rm keV}) \lesssim -1.7$) are required. The softer AGN ionizing spectra indicate that, if any of these `N-enhanced' galaxies are AGN-dominated, their black hole masses and Eddington accretion rates differ from the $z<4$ AGN sample \citep{thomas_physically_2016}. Alternatively, heavy dust obscuration near the AGN could also soften the AGN ionizing spectrum. Moreover, only `N-enhanced' \ion{H}{2} and AGN models successfully predict sufficiently high \ion{N}{3}]/\ion{C}{3}] and \ion{N}{4}]/\ion{C}{3}] ratios to match the high-$z$ observations, as shown in Figure~\ref{fig:3} (b) and (c). The predictions from `N-normal' models fall short by approximately 1.5 dex. Although shock models can produce modestly elevated \ion{N}{3}]/\ion{C}{3}] ratios that are about 0.5 dex higher than \ion{H}{2} and AGN models with the same nitrogen abundance \citep{flury_new_2024}, they can not fully account for the strength of the observed nitrogen lines in these galaxies. }

\begin{figure*}[htb]
\epsscale{1.2}
\plotone{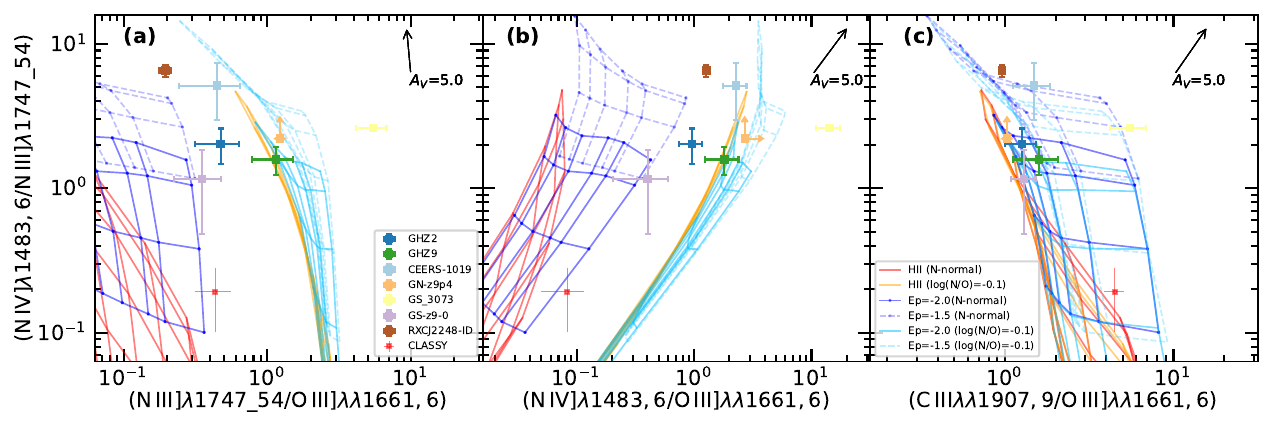}
\caption{Comparison of \ion{H}{2} and AGN photoionization models with observations on NIV/NIII-based UV diagnostic diagrams. Model grid parameter ranges and symbols are the same as in Figure~\ref{fig:1}, except that HII models now have gas pressures of $\log \rm(P/k)=9.2$, and AGN models with $\log (E_{\text{peak}}/\text{keV})=-2.0$ and $\log (E_{\text{peak}}/\text{keV})=-1.5$ are shown in solid and dashed model grids, respectively. {\color{black}These UV diagrams are effective in identifying elevated N/O ratios and constraining gas-phase metallicity.}
 \label{fig:4}}
\end{figure*} 

The \ion{N}{4}]/\ion{N}{3}]-\ion{N}{3}]/\ion{O}{3}] and \ion{N}{4}]/\ion{N}{3}]-\ion{N}{4}]/\ion{O}{3}] UV diagnostic diagrams {\color{black}shown in Figure~\ref{fig:4}} are most effective in identifying nitrogen-enhanced galaxies, because the \ion{N}{3}]/\ion{O}{3}] and \ion{N}{4}]/O~III ratios are very sensitive to the N/O ratio. {\color{black} Additionally, these diagrams exclude \ion{C}{4} and \ion{He}{2} and thus are not affected by stellar wind, ISM absorption, or resonant scattering.} Figure~\ref{fig:4}(a) and (b) show that GS 3093 has the highest N/O ratio, which is exceeding log(N/O)=$-0.1$. GHZ9 and GN-z9p4 have the second highest N/O ratios, which are around log(N/O)=$-0.1$. GHZ2, CEERS-1019, RXCJ2248-ID, and GS-z9-0 have elevated N/O compared to the local values, but not as high as log(N/O)=$-0.1$.  These inferred nitrogen abundances agree with literature values listed in Table~\ref{tab:1}.

The \ion{N}{4}]/\ion{N}{3}]-\ion{C}{3}]/\ion{O}{3}] UV diagnostic diagram could be used to constrain the gas-phase metallicity of the host galaxy, if the carbon abundance follows the local carbon scaling relation. As shown in Figure~\ref{fig:4}(c), galaxies with higher gas-phase metallicity have higher \ion{C}{3}]/\ion{O}{3}] ratio, and the \ion{C}{3}]/\ion{O}{3}] ratio is consistent across AGN and \ion{H}{2} models at fixed $12+\log(\mathrm{O/H})$ regardless of changes in nitrogen abundance. Therefore, we could infer from Figure~\ref{fig:4}(c) that GS 3073 has the highest gas-phase metallicity of $12+\log(\mathrm{O/H})\gtrsim8.0$  among our high-$z$ galaxy sample, and all other galaxies have lower gas-phase metallicities of $12+\log(\mathrm{O/H})\lesssim8.0$. 

{\color{black} Although effective for identifying nitrogen anomalies and constraining metallicity,} the \ion{N}{4}]/\ion{N}{3}]-based UV diagrams have limited power to isolate AGN contamination. The \ion{N}{4}]/\ion{N}{3}] ratio is not only sensitive to the hardness of the ionizing spectrum, but also varies strongly with the ionization parameter and becomes pressure-sensitive at $\log(P/k)\gtrsim7.8$. As illustrated in Figure~\ref{fig:4}, the high-$z$ nitrogen-enhanced galaxies can be explained either by \ion{H}{2} models with extremely high gas pressure ($\log(P/k)\gtrsim9.2$) and high ionization parameter ($\log(U) \gtrsim -1.0$), or by AGN models with moderate gas pressure ($\log(P/k)\gtrsim7.0$) and less extreme ionization parameter ($\log(U)\gtrsim-2.0$ for AGN with $\log(E_{\text{peak}}/\text{keV})\gtrsim-1.5$). Additional diagnostic diagrams are required to further constrain the excitation mechanisms in these galaxies.

\begin{figure*}[htb]
\epsscale{1.0}
\plotone{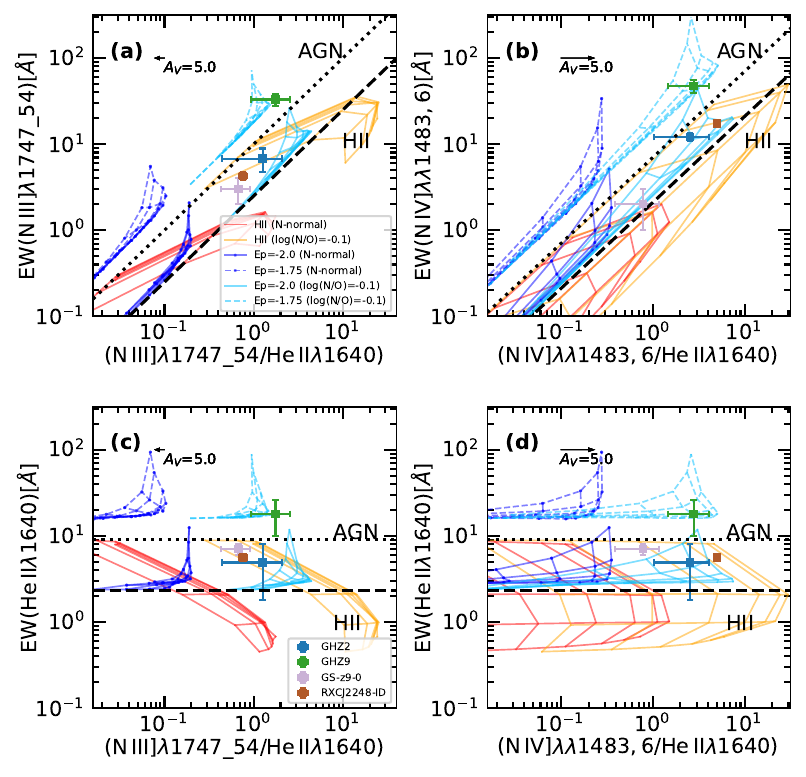}
\caption{{\color{black}The most effective UV diagnostic diagrams to separate \ion{H}{2} regions and AGN independent of varying nitrogen abundance.} AGN models with $\log (E_{\text{peak}}/\text{keV})=-2.0$ and $\log (E_{\text{peak}}/\text{keV})=-1.75$ are shown in solid and dashed model grids. The gas metallicity range is now within $12+\log(\rm O/H)\leq8.43$ for all models, corresponding to the metallicity range of the $z>5$ nitrogen-enhanced galaxy listed in Table~\ref{tab:1}. Other model grid parameter ranges are the same as in Figure~\ref{fig:1}, except that HII models now have gas pressures of $\log \rm(P/k)=9.2$. In each panel, the black dashed line represents the division line between star-forming galaxies (\ion{H}{2}) and ambiguous galaxies, and the black dotted line represents the division line between AGN-dominated galaxies and ambiguous galaxies, respectively. The division lines are defined and justified in the main text.
 \label{fig:5}}
\end{figure*} 

{\color{black}Equivalent width (EW)-based UV diagnostic diagrams, previously investigated by \citet{nakajima_vimos_2018,hirschmann_synthetic_2019}, are the most effective UV diagrams to identify AGN contamination from star formation, as shown in Figure~\ref{fig:5}.} AGN models with varying $E_{\text{peak}}$ values are clearly separated in these EW-based diagnostic diagrams. This separation arises because both EWs and emission line ratios involving \ion{He}{2} are sensitive to the hardness of the ionizing radiation field. Harder radiation fields produce stronger \ion{He}{2} emission and higher EWs, causing AGN models with higher $E_{\text{peak}}$ to shift toward the upper-left corner of Figure~\ref{fig:5}(a) and (b).

AGN models with $\log(E_{\text{peak}}/{\rm keV})\lesssim-1.9$ overlap with metal-poor ($12+\log(\rm O/H)\leq8.0$) \ion{H}{2} models on these UV diagnostic diagrams, indicating a similarity in the hardness of their radiation fields. This similarity is not surprising because low metallicity stellar populations tend to have higher effective temperatures, resulting in harder ionizing spectra \citep[e.g.][]{choi_evolution_2017,grasha_stromlo_2021}. Compared to low-metallicity stellar populations, star-forming galaxies with moderate or high metallicity ($12+\log(\rm O/H)\geq8.0$) have softer radiation fields, shifting them toward the lower-right corner of Figure~\ref{fig:5}(a) and (b) and distinguishing them from AGN emissions.

{\color{black}EW(\ion{He}{2})-based UV diagnostic diagrams have a similar ability to separate strong AGNs from pure star-forming galaxies, as shown in Figure~\ref{fig:5}(c) and (d). This is because the EW of \ion{He}{2} is mostly sensitive to the hardness of the radiation field and has only a weak dependence on the ionization parameter and gas pressure at $\log(U)\lesssim-1.5$. In addition, the EW(\ion{He}{2}) is insensitive to nitrogen abundance, making these diagrams robust discriminators.

Considering only star-forming galaxies and AGN, we can divide galaxies into three categories (Star-forming galaxies, ambiguous galaxies, and AGNs) based on their locations on these EW-based UV diagnostic diagrams. The division lines between different categories are shown in Figure~\ref{fig:5} using dashed and dotted lines and described below.}

(1) Star-forming Galaxies: Galaxies in the following regions are dominated by star formation with moderate or high metallicity ($12+\log(\rm O/H)\geq8.0$).
{\small
\begin{align*}
\rm{EW}(\mathrm{He\,II})(\text{\AA}) &\leq 2.3\\
\mathrm{EW}(\mathrm{N\,III]})(\text{\AA}) &\leq 2.5 \times \frac{\mathrm{Flux}(\mathrm{N\,III]})}{\mathrm{Flux}(\mathrm{He\,II})} \\
\mathrm{EW}(\mathrm{N\,IV]})(\text{\AA}) &\leq 2.1 \times \frac{\mathrm{Flux}(\mathrm{N\,IV]})}{\mathrm{Flux}(\mathrm{He\,II})} 
\end{align*}}
{\color{black}
(2) AGNs: Galaxies in the following regions are dominated by AGNs that have $\log(E_{\text{peak}}/{\rm keV})\gtrsim-1.8$.
{\small
\begin{align*}
\rm{EW}(He~II]) (\text{\AA})& \geq 9 \\
\rm{EW}(N~III]) (\text{\AA})&\geq 10 \times \frac{\rm Flux(N~III])}{\rm Flux(He~II)}\\
\rm{EW}(N~IV]) (\text{\AA})&\geq 7 \times \frac{\rm Flux(N~IV])}{\rm Flux(He~II)}
\end{align*}}
(3) Ambiguous Galaxies: Galaxies in the following regions are either metal-poor star-forming galaxies or AGNs that have a soft ionizing spectrum ($\log(E_{\text{peak}}/{\rm keV})\lesssim-1.9$). Alternatively, these galaxies could also be excited by a comparable mixture of star formation and AGN activity. 
{\small
\begin{align*}
9 \geq \rm{EW}(He~II]) (\text{\AA}) & \geq 2.3 \\
10 \times \frac{\rm Flux(N~III])}{\rm Flux(He~II)}\geq \rm{EW}(N~III]) (\text{\AA}) &\geq 2.5 \times \frac{\rm Flux(N~III])}{\rm Flux(He~II)} \\
7 \times \frac{\rm Flux(N~IV])}{\rm Flux(He~II)} \geq \rm{EW}(N~IV]) (\text{\AA})&\geq 2.1 \times \frac{\rm Flux(N~IV])}{\rm Flux(He~II)}
\end{align*}}
}
{\color{black}Figure~\ref{fig:5} shows that GHZ9 is located in the `AGN' region, indicating AGN-dominated excitation. The best-fitting AGN model to GHZ9 has a relatively soft ionizing spectrum ($\log(E_{\text{peak}}/{\rm keV}) = -1.75$), high ionization parameter ($\log(U) \approx -1.0$), and high gas pressure ($\log(P/k) > 8.4$). In contrast, GHZ2, GS-z9-0, and RXCJ2248-ID lie in the `ambiguous galaxies' region, indicating that they could be star-forming galaxies, weak AGNs, or have a composite (stellar and AGN) excitation.}

\begin{figure}[htb]
\epsscale{1.1}
\plotone{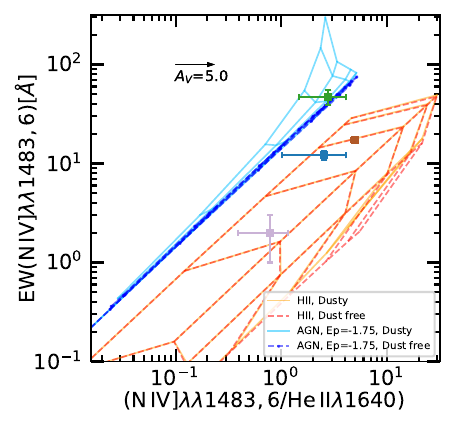}
\caption{{\color{black} Comparison of dusty (solid grids) and dust-free (dashed grids) \ion{H}{2} and AGN photoionization models. Model parameter ranges match those in Figure~\ref{fig:5}. Dust-free models do not alter the \ion{H}{2} and AGN division lines derived from dusty models.} \label{fig:8}}
\end{figure}

{\color{black}We also test whether the \ion{H}{2}--AGN division lines derived from dusty models remain valid for galaxies with little or no dust. Across all UV diagnostic diagrams, the dusty and dust-free grids agree closely at low metallicity ($12+\log(\rm O/H)\lesssim8.4$), with line ratio offsets $\Delta(\log R)\lesssim0.1\,$dex. As an illustration, Figure~\ref{fig:8} shows the comparison of dusty and dust-free \ion{H}{2} and AGN models on the EW(\ion{N}{4}])-\ion{N}{4}]/\ion{He}{2} diagnostic diagram. At $12+\log(\rm O/H)\lesssim8.2$, the location of dusty and dust-free \ion{H}{2} and AGN models are mostly identical, indicating that division lines derived from dusty models are applicable to dust-free galaxies. The largest differences arise only at higher metallicity ($12+\log(\rm O/H)\geq8.4$), high ionization parameter ($\log(U)\geq-1.5$), and high gas pressure ($\log(P/k)\approx9.2$), where dust photoelectric heating raises the electron temperature and yields slightly larger EWs and stronger \ion{He}{2}. These differences do not affect our conclusions.}

\vspace{-2em} 

\section{Constraining Excitation Sources and Gas Properties}\label{sec:5}

Although UV diagnostic diagrams are valuable tools for distinguishing hard AGN excitations from star-forming activities, their ability to constrain gas metallicity, gas pressure, and ionization parameter is limited due to the parameter degeneracies within the two-dimensional diagnostic space. They also have a limited ability to distinguish between soft AGN excitations and hard star-forming excitations, which often occupy overlapping regions in the UV diagnostic diagrams. Additionally, UV diagnostic diagrams only incorporate bright UV emission lines{\color{black}, which are most sensitive to high-energy photons and preferentially trace high-ionization gas, limiting leverage on lower-ionization gas.} As a result, a best-fit model to a galaxy spectrum in the UV diagnostic diagrams is not guaranteed to provide a good fit at optical wavelengths. To further constrain the excitation source and gas properties, a comprehensive cross-match between the model and all observed emission line properties of a galaxy is necessary {\color{black}\citep{olivier_characterizing_2022,fernandez_new_2022,marconi_homerun_2024}}. 

\begin{figure*}[htb]
\epsscale{1.0}
\plotone{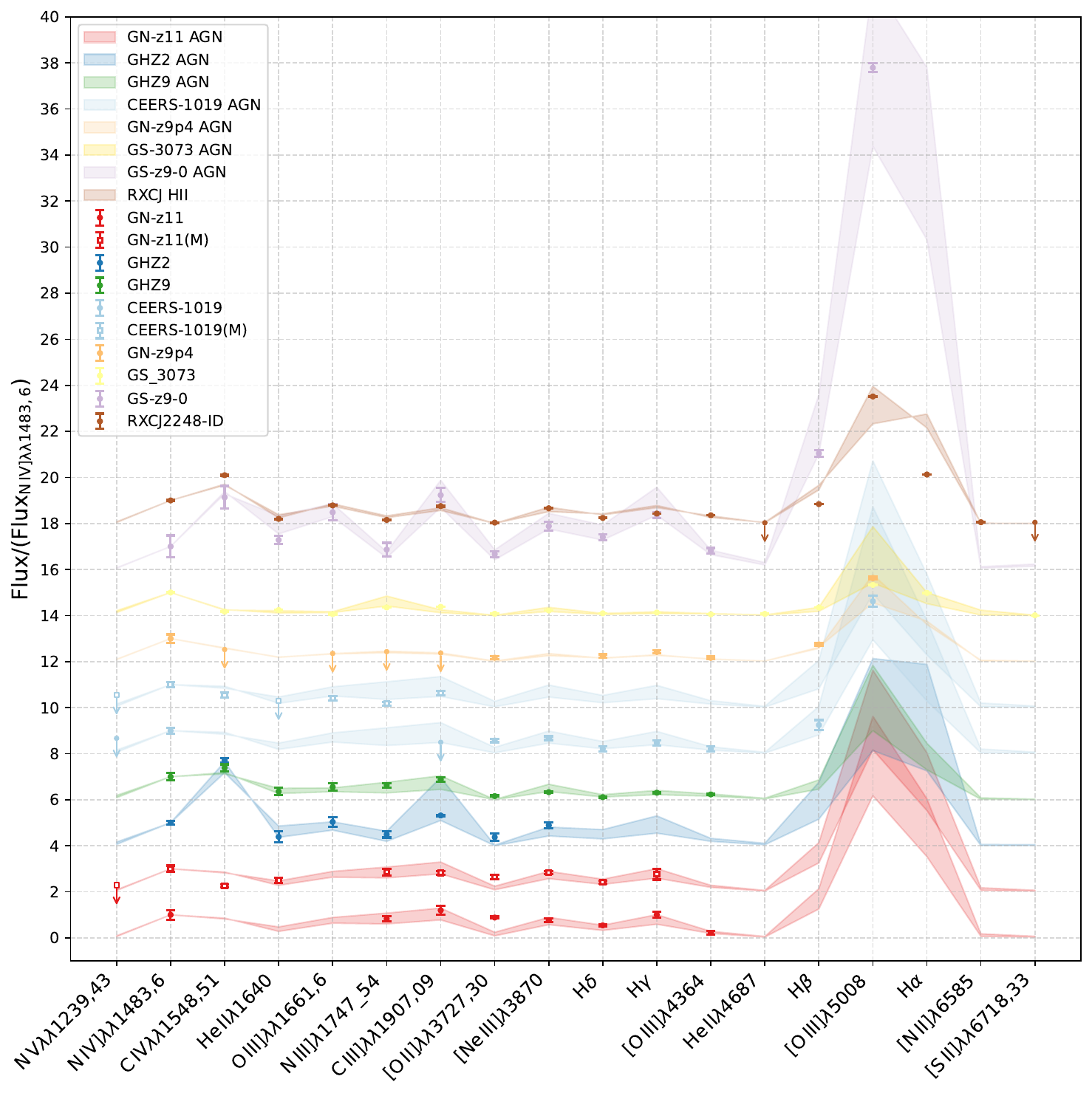}
\caption{Comparison of best-fit models with observations on UV and optical emission line fluxes. Observational data are depicted as data points with error bars, with each color representing a different galaxy. The PRISM spectra of GN-z11 and CEERS-1019 are measured by two different references. The solid circle and open square represent measurements from \citet{bunker_jades_2023} and \citet{maiolino_small_2024} for GN-z11, \citet{larson_ceers_2023} and \citet{marques-chaves_extreme_2024} for CEERS-1019, respectively. For each observation, the y-axis displays the emission line fluxes normalized to the \ion{N}{4}] fluxes. To enhance visual clarity, the data for each galaxy is vertically offset by 2. The shaded regions correspond to the range of best-fit model predictions for each galaxy, with boundaries defined by two models (model 1 and model 2) that share the same excitation source and nitrogen abundance but exhibit slight differences in gas metallicity, ionization parameter, or gas pressure.
 \label{fig:6}}
\end{figure*} 

In this section, we explore the tailored photoionization models for each of the eight $z\gtrsim5$ nitrogen-enhanced galaxies in our sample. Our goal is to identify best-fit models that simultaneously reproduce the observed UV emission-line fluxes, optical emission-line fluxes, and UV equivalent width measurements. In this comparison, all emission line fluxes are corrected for dust extinction using the $A_V$ values listed in Table~\ref{tab:1}, assuming the Milky Way value $R_V=3.1$, and applying the dust extinction curve in \citet{cardelli_relationship_1989}. This approach enables robust constraints on key physical properties of each galaxy, including the dominant excitation source, gas-phase metallicity, nitrogen abundance, ionization parameter, and gas pressure.

To initialize the model parameters, we adopt literature-reported estimations for nitrogen abundance, metallicity, and gas pressure (using the ideal gas law, $P=nkT$). Initial values for the ionization parameter and excitation mechanism are inferred from the locations of galaxies in Figures~\ref{fig:1}$-$~\ref{fig:5}. For GHZ9 and GS 3073, we explore only AGN models.  For the remaining six galaxies, we consider both \ion{H}{2} model and AGN models with $\log(E_{\text{peak}}/{\rm keV})\lesssim-1.85$.

The best-fit model for each galaxy is explored by iteratively adjusting the gas-phase metallicity (in 0.05 dex steps), nitrogen abundance (0.1 dex), ionization parameter (0.2 dex), and gas pressure (0.6 dex) around the initial values. For AGN models, $\log(E_{\text{peak}}/{\rm keV})$ is also varied in 0.05 dex steps. For GHZ2 and GHZ9, we also adjust the carbon abundance because their exceptionally large \ion{C}{4} equivalent widths ($>40$ \AA) cannot be reproduced along with other observational properties using the local carbon scaling relation in \citet{2017MNRAS.466.4403N}. The model that exhibits the best overall agreement with the observed UV and optical emission-line fluxes, as well as UV equivalent widths, is selected as the best-fit model based on visual inspection. For a larger sample, a more sophisticated approach, such as Bayesian inference \citep{thomas_interrogating_2018,li_massmetallicity_2024} or Monte Carlo sampling, is recommended.

\begin{figure*}[htb]
\epsscale{1.1}
\plotone{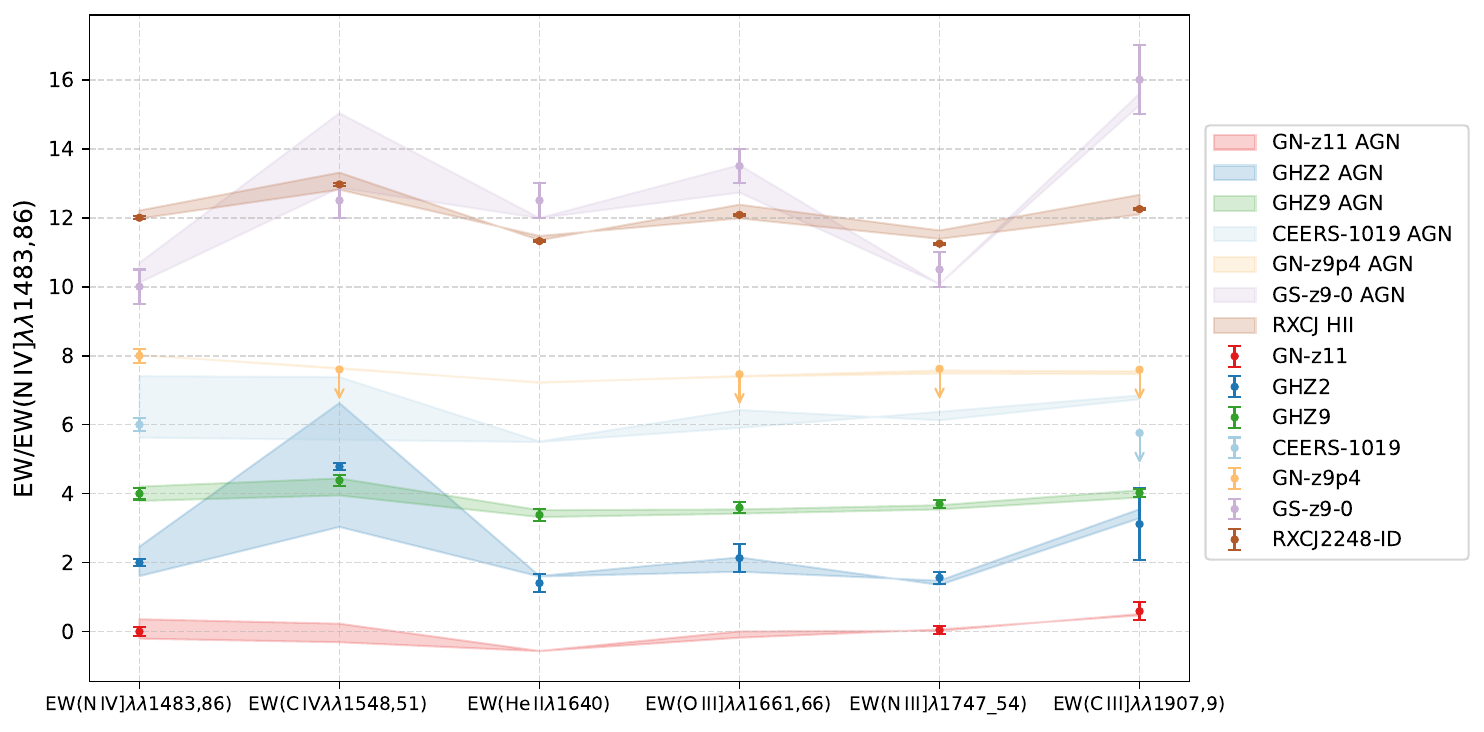}
\caption{Comparison of best-fit model predictions with observed rest-frame UV equivalent width (EW) measurements. Observational data are depicted as data points with error bars, with each color representing a different galaxy. The EWs for GN-z11 and CEERS-1019 are measured by \citet{bunker_jades_2023} and \citet{larson_ceers_2023}, respectively. For each galaxy, the y-axis displays the EW of emission lines normalized to the EW of \ion{N}{4}]. To enhance visual clarity, the data for each galaxy is vertically offset by 2. The shaded regions correspond to the range of best-fit model predictions for each galaxy, with boundaries defined by two models (model 1 and model 2) that share the same excitation source and nitrogen abundance but exhibit slight differences in gas metallicity, ionization parameter, or gas pressure.
 \label{fig:7}}
\end{figure*}

{\color{black} We present the comparison between our best-fit model predictions and the observed UV and optical emission-line fluxes in Figure~\ref{fig:6}, and the corresponding comparison for UV EW measurements in Figure~\ref{fig:7}}. For each galaxy, the model prediction is represented by a shaded region bounded by two models that share the same excitation source and nitrogen abundance but differ slightly in gas metallicity, ionization parameter, or gas pressure. Instead of fixing these gas properties at single values, we allow for small variations to account for spatial inhomogeneities within the galaxy regions encompassed by the PRISM spectroscopic aperture. This strategy offers a more realistic representation of the internal conditions within each galaxy. 

Overall, the best-fit models reproduce $80\%-100\%$ of the observed emission-line fluxes and EW measurements within the observational uncertainties. As shown in Figures~\ref{fig:6} and~\ref{fig:7}, most data points fall within the predicted shaded regions. Deviations within $\sim0.25$ in relative flux ratios can generally be attributed to uncertainties in the model predictions, which are typically $\lesssim0.1$ dex in logarithmic space \citep{kewley_theoretical_2001}, equivalent to $\lesssim0.2-0.3$ in linear scale. 

Discrepancies larger than $\sim0.3$ may arise from various factors. For instance, mismatches in \ion{C}{4} fluxes for GN-z11, CEERS-1019, and RXCJ2248-ID, as well as the discrepancy in EW(CIII) for CEERS-1019, may reflect differences between the assumed and actual carbon abundances in these galaxies, {\color{black}because simply raising the model gas pressure can not resolve these issues without affecting the overall fits.} The systematically lower hydrogen line fluxes observed in RXCJ2248-ID may suggest that a minor broad component in the Balmer lines was not accounted for in the flux measurements. Indeed, \citet{topping_metal-poor_2024} reported the presence of a broad H$\alpha$ component in RXCJ2248-ID, contributing approximately 22\% of the total flux with a FWHM of 607 km/s. However, they also noted that corresponding broad components in H$\beta$ and other Balmer lines, if present at similar flux ratios, would fall below the detection threshold in their observations. {\color{black}The mismatch in the optical [O~II] emission line flux of GN-z11 may result from a missing soft radiation field, likely \ion{H}{2} regions in the host galaxy, which contribute additional low-ionization emission.} 

Since our main focus is the nitrogen abundance and the primary excitation source of these high-$z$ galaxies, we limit our exploration to single-source photoionization models. As shown in Figures~\ref{fig:6} and ~\ref{fig:7}, single-source photoionization models with small variations in gas-phase metallicity, ionization parameter, and gas pressure are already remarkably successful in reproducing the majority of the observed emission line ratios and EW measurements. The parameters of the best-fit models for each galaxy are summarized in Table~\ref{tab:2}. 

\setlength{\tabcolsep}{2pt} 
\begin{deluxetable*}{c|c|c|c|c|c|c|c|c}[hbt]
\small
\centering
\tablewidth{0pc}
\tablecaption{Parameters of Best-fit Models for Nitrogen Enhanced Galaxies \label{tab:2}}
\tablenum{2}
\tablehead{
\colhead{Galaxy} &
\colhead{Excitation Source} &
\colhead{$\log$(N/O)} &
\colhead{$\log$(C/O)\tablenotemark{\scriptsize a}} &
\colhead{12+$\log(\rm O/H)$} &
\colhead{$\log (P/k)$} &
\colhead{\color{black}$\log (n/ \rm cm^{3})$\tablenotemark{\scriptsize c}} &
\colhead{$\log (U)$} &
\colhead{Match Fraction\tablenotemark{\dag}}
}
\startdata
{GN-z11} & AGN, $\log(E_{\text{peak}}/{\rm keV})= -1.95$ & $-0.2$ & $--$ & $7.5$ & $7.8$ & 3.4-3.8 & $(-1.5,-1.2)$\tablenotemark{\scriptsize b} & 80\%-91\% \\
{GHZ2} 	 & AGN, $\log(E_{\text{peak}}/{\rm keV})= -1.9$ & $-0.4$ & $-0.4$ & $7.3$ & $9.2$ & 4.8-5.2 & $(-1.5,-0.8)$\tablenotemark{\scriptsize b}  & 93\% \\
{GHZ9} 	 & AGN, $\log(E_{\text{peak}}/{\rm keV})= -1.75$ & $-0.1$ & $-0.6$ & $(7.8,7.9)$\tablenotemark{\scriptsize b} & $9.8$ & 5.4-5.8 & $(-1.7,-1.2)$\tablenotemark{\scriptsize b} & 95\% \\
CEERS-1019  & AGN, $\log(E_{\text{peak}}/{\rm keV})= -1.95$ & $-0.2$ & $--$ & $7.6$ & $7.8$ & 3.4-3.8 & $(-1.5,-0.8)$\tablenotemark{\scriptsize b} & 86\%-92\% \\
GN-z9p4	 & AGN, $\log(E_{\text{peak}}/{\rm keV})= -1.9$ & $0.0$ & $--$ & $(7.4,7.7)$\tablenotemark{\scriptsize b}  & $(9.2,8.4)$\tablenotemark{\scriptsize b} & 4.4-5.2 & $-1.2$ & 100\% \\
GS 3073	 & AGN, $\log(E_{\text{peak}}/{\rm keV})= -1.75$ & $0.4$ & $--$ & $(8.2,8.0)$\tablenotemark{\scriptsize b} & $(9.2,9.8)$\tablenotemark{\scriptsize b} & 4.8-5.8 & $(-1.5,-1.0)$\tablenotemark{\scriptsize b} & 100\% \\
GS-z9-0	 & AGN, $\log(E_{\text{peak}}/{\rm keV})= -1.9$ & $-0.8$ & $--$ & $7.35$ & $7.0$ & 2.6-3.0 & $(-1.6,-1.4)$ \tablenotemark{\scriptsize b} & 95\% \\
RXCJ2248-ID	& \ion{H}{2} & $-0.4$ & $--$ & $(7.2,7.4)$\tablenotemark{\scriptsize b} & $9.8$ & 5.4-5.8 & $-1.0$ & 83\% \\
\enddata
\tablenotetext{\scriptsize a}{$--$ represents that carbon abundance follows the scaling relations in \citet{2017MNRAS.466.4403N} and is not manually adjusted.}
\tablenotetext{\scriptsize b}{Parenthetical pairs (A, B) denote the values adopted in model 1 and model 2, respectively.}
\tablenotetext{\scriptsize c}{\color{black}Gas density is inferred from the gas pressure via $P=nkT$, adopting $T\approx(1\text{–}2.5)\times10^4$ K. These values are provided for comparison with direct measurements only. Density and temperature vary across the nebula, and detailed treatments are required to recover the local $n$ and $T$ in specific zones.}
\tablenotetext{\dag}{Fraction of observed quantities (fluxes and EWs) that are successfully reproduced by the best-fit models.}
\end{deluxetable*}
\vspace{-2.5em} 

One notable feature of the best-fit models is that 7 out of the 8 galaxies in our sample are best described by AGN models. Among them, two galaxies (GHZ9 and GS 3073) require moderately hard AGN radiation fields with $\log(E_{\text{peak}}/{\rm keV})= -1.75$, while the other five galaxies (GN-z11, GHZ2, CEERS-1019, GN-z9p4, and GS-z9-0) require softer AGN radiation fields ($\log(E_{\text{peak}}/{\rm keV})\approx-1.9$). Only RXCJ2248-ID is best fitted by a metal-poor \ion{H}{2} model with high gas pressure, {\color{black}in agreement with \citet{topping_metal-poor_2024}.}  

Another remarkable feature is that all best-fit models have moderate to high gas pressure $7.0\leq\log (P/k)\leq9.8$. {\color{black}This gas pressure range corresponds to gas density range of $2.6\leq \log n(\rm cm^{-3})\leq5.8$ for $T\approx(1-2.5)\times10^4\,$K}. These densities are consistent with the direct electron density measurements in CEERS-1019, GS 3073, GS-z9-0, and RXCJ2248-ID, and provide constraints for GHZ2, GHZ9, and GN-z9p4 where direct measurements of electron density are unavailable. 

Regarding gas-phase metallicity and nitrogen enrichment, our best-fit models are generally consistent with values reported in the literature to within $\sim0.1\,$dex. This confirms that these galaxies are metal-poor and exhibit elevated N/O ratios. Importantly, this consistency also suggests that the presence of AGNs does not significantly affect the inferred N/O ratios. A notable exception is GN-z9p4, for which we derive a significantly higher N/O ratio ($\log(\rm N/O)=0$) compared to the value of $\log(\mathrm{N}/\mathrm{O}) = -0.6$ reported by \citet{schaerer_discovery_2024}. {\color{black}This discrepancy resulted from the different choices of electron densities. We discuss this further in Section~\ref{sec:6.2}.} 

Additionally, we find that GHZ2 and GHZ9 require an elevated carbon-to-oxygen (C/O) ratio to reproduce their carbon emission line features, which are not suggested in \citet{castellano_jwst_2024} and \citet{napolitano_dual_2024}. {\color{black}Using the non-linear carbon abundance scaling relation from \citet{2017MNRAS.466.4403N}, our photoionization models cannot simultaneously reproduce the carbon lines and the rest of the UV and optical features, even at extreme conditions ($\log (P/k)=9.8$, $\log(U)\gtrsim-1.0$). Allowing a modest C/O enhancement above the local carbon scaling relation resolves these mismatches while preserving the global fit to the other lines.}

We also see evidence of changing gas density in different nebular zones in GN-z11. Our best-fit model suggests a gas pressure of $\log (P/k) = 7.8$, which is consistent with the value inferred from the optical [O~II] doublet ratio ([O~II] $\lambda3729/\lambda3727 = 0.59$) reported in \citet{maiolino_jades_2024} (e.g., using the AGN gas pressure diagnostics in \citet{zhu_theoretical_2024}). However, the electron densities derived from the \ion{N}{4}] doublet ratio and the \ion{N}{3}] quintuplet are $n_{\rm e}\gtrsim10^6\rm \,cm^{-3}$ or even higher \citep{maiolino_jades_2024}, which is more than two orders of magnitude higher than the electron density in our best-fit model. {\color{black}Similar gas density variation is also observed in GS 3073 \citep{ji_ga-nifs_2024}, which reflects the density fluctuation within photoionization regions. We discuss this further in Section~\ref{sec:6.2}.}

In addition to the high gas pressure, these best-fit models also exhibit higher ionization parameters $\log(U)\gtrsim-2.0$ than the local optical-selected emission line galaxies $\log(U)\lesssim-2.0$. Given that their gas pressures are also higher than the local galaxies, the enhanced ionization parameters imply that ionized nebulae are located closer to their excitation sources. This interpretation follows from the definition of the ionization parameter, $U \sim Q_{\rm ion} / (4\pi r^2 n_{\rm H} c)$, where $Q_{\rm ion}$ is the ionizing photon rate, $r$ is the distance to the source, $n_{\rm H}$ is the hydrogen number density, and $c$ is the speed of light \citep[see][for a review]{kewley_understanding_2019}. The inferred compactness of the ionized regions is consistent with the compact morphologies of these galaxies ($R_e \lesssim 120\,\rm{pc}$) reported in the corresponding references.

\section{Discussion}

\subsection{Excitation Source: Star Formation or AGN?}

The AGN nature for GHZ9 and GS 3073 is consistent with previous findings in the literature. X-ray emission spatially coincident with GHZ9 was first reported by \citet{kovacs_candidate_2024} and subsequently confirmed by \citet{napolitano_dual_2024}. X-ray observations provide compelling evidence that GHZ9 hosts an actively accreting supermassive black hole (SMBH). Assuming a photon index of $\Gamma = 2.3$, \citet{napolitano_dual_2024} estimates the black hole mass of GHZ9 to be $M_{\rm BH} = 1.6 \times 10^8\,M_{\odot}$. GS 3073 is reported as a Type 1.8 AGN in \citet{ubler_ga-nifs_2023}, where prominent broad emission lines are found in the rest-frame optical spectrum observed by JWST/NIRSpec IFU. The luminosity and line width of the broad H$\alpha$ emission suggest a central BH mass of $\log(M_{\rm BH}/M_{\odot})\approx8.2$ with the Eddington ratio of $f_{\rm Edd}\approx0.1-1.6$ in GS 3073 \citep{ubler_ga-nifs_2023}. 

We can infer the Eddington ratio from $E_{\text{peak}}$ and $M_{\rm BH}$ values using the empirical relation in Figure 3 of \citet{thomas_physically_2016}. For GHZ9 and GS 3073 whose $\log(E_{\text{peak}}/{\rm keV})= -1.75$ and $\log(M_{\rm BH}/M_{\odot})\approx8.2$, their Eddington ratios are $f_{\rm Edd}\approx0.3$. This is consistent with the Eddington ratio estimation for GS 3073 reported in \citet{ubler_ga-nifs_2023}.

{\color{black} The six remaining galaxies, GN-z11, GHZ2, CEERS-1019, GS-z9-0, GN-z9p4, and RXCJ2248-ID lie in the \ion{H}{2}-AGN overlap region on the UV diagnostic diagrams, making these diagrams alone unable to identify their dominant excitation sources. This overlap reflects a physical degeneracy: spectra powered by metal-poor stellar populations (12+$\log(\mathrm{O}/\mathrm{H}) \lesssim 7.7$) are hard enough to mimic those powered by the softest AGN radiation fields ($\log(E_{\text{peak}}/{\rm keV})\lesssim-1.9$). For galaxies with higher metallicity (12+$\log(\mathrm{O}/\mathrm{H})\gtrsim 7.7$) within the overlap region, a purely stellar origin is disfavored because the stellar ionizing spectrum softens with increasing metallicity. However, all six galaxies are metal-poor, urging the need to include all observed emission-line fluxes and EW measurements in the model fitting to break this degeneracy.

From the comprehensive fits in Section 5, five galaxies (GN-z11, GHZ2, CEERS-1019, GS-z9-0, GN-z9p4) are best described by AGN photoionization models. Although a partial stellar contribution cannot be excluded, pure \ion{H}{2} models fail to reproduce the observed fluxes and EWs simultaneously, indicating AGN-dominated excitation. In the thin-disk AGN accretion regime \citep{done_intrinsic_2012,thomas_physically_2016}, these soft AGN radiation fields ($\log(E_{\text{peak}}/{\rm keV})\lesssim-1.9$) correspond to intermediate BHs with $6.0\lesssim\log(M_{\rm BH}/M_{\odot})\lesssim8.0$ accreting at low Eddington ratios ($0.001\lesssim f_{\rm Edd}\lesssim0.03$), or SMBHs with $\log(M_{\rm BH}/M_{\odot})\gtrsim8.0$ accreting at moderate to high rates ($0.1\lesssim f_{\rm Edd}\lesssim1$).

Independent evidence generally supports these classifications. GN-z11 shows extremely high electron density of $n_{\rm e}\gtrsim10^6\,\rm cm^{-3}$ from the UV \ion{N}{4}] doublet ratio and the \ion{N}{3}] quintuplet that are consistent with gas in AGN BLRs \citep{maiolino_small_2024}, despite showing no broad emission lines in the rest-frame optical spectrum \citep{alvarez-marquez_insight_2025}. CEERS-1019 exhibits a $\sim2.5\sigma$ detection of broad H$\beta$ emission line \citep{larson_ceers_2023,marques-chaves_extreme_2024}. GS-z9-0 shows a $\sim3\sigma$ detection of high-ionization line [Ne~V]$\lambda3426$ \citet{curti_jades_2024}. For GHZ2 and GN-z9p4, neither [Ne~V]$\lambda3426$ emission nor broad components are detected at current depth \citep{castellano_jwst_2024,schaerer_discovery_2024}. This could result from sensitivity limits or indicate that they are intrinsically obscured weak AGNs. 

Only RXCJ2248-ID is best described by \ion{H}{2} models, leaving it the only pure star-forming galaxy in our `N-enhanced' high-redshift galaxy sample. The best-fit \ion{H}{2} region model involves a metal-poor stellar population (12+$\log(\rm O/H)\approx7.3$) embedded in the dense ($n_{\rm e}\gtrsim 10^5\,\rm{cm}^{-3}$) and compact ($\log(U)=-1$) ISM. This classification is consistent with \citet{topping_metal-poor_2024}, who find no evidence of {\color{black}[Ne~V]$\lambda3426$} or broad line features in its spectra.}

\subsection{What Affect the Measurements of N/O Ratio?} \label{sec:6.2}

{\color{black}
N/O measurements could be strongly affected by the assumption of electron density. An example is GN-z9p4, for which our best-fit model yields $\log(\rm N/O)=0$, much higher than the $\log(\mathrm{N}/\mathrm{O})=-0.6$ reported in \citet{schaerer_discovery_2024}. This discrepancy is mainly due to differences in the electron density $n_{\rm e}$. In the absence of a direct $n_{\rm e}$ measurement, \citet{schaerer_discovery_2024} assumed the low density limit $n_{\rm e} = 10^2\,\rm{cm}^{-3}$ in the abundance estimation and inferred $T_e(\rm OIII)\approx2.3\times10^4\rm\,K$. By contrast, our best-fit model yields gas pressures $\log(P/k)\approx8.4-9.2$, implying $n_{\rm e}\approx10^4-10^5\,\rm{cm}^{-3}$ and $T_e\approx(1-1.5)\times10^4\rm\,K$. \citet{schaerer_discovery_2024} noted that at very high electron density ($n_{\rm e}\gtrsim10^5\,\rm cm^{-3}$), the inferred $T_e$ would be lower and lead to a higher O/H and even more higher N/O and C/O ratios. This is consistent with our results. 

However, electron density varies across the photoionization region, further complicating N/O measurements. Within our sample, electron density variations are observed in both GS 3073 and GN-z11. In GS 3073, its optical [\ion{S}{2}] and [\ion{Ar}{4}] doublet ratios yield $n_{\rm e}\sim 10^3\,\rm{cm}^{-3}$ \citep{ubler_ga-nifs_2023}, whereas the UV \ion{O}{3}] doublet indicates a significantly higher density of $n_{\rm e}\sim 10^5\,\rm{cm}^{-3}$ \citep{ji_ga-nifs_2024}. In GN-z11, its optical [\ion{O}{2}] ratio and our best-fit model both reported a gas pressure of $\log (P/k) = 7.8$, whereas the UV \ion{N}{4}] doublet ratio and the \ion{N}{3}] quintuplet yield $n_{\rm e}\gtrsim10^6\rm \,cm^{-3}$ or even higher \citep{maiolino_small_2024}. 

Variations in gas density traced by different ionic species are expected, as these emission lines originate from distinct zones within the nebula. High-ionization lines like \ion{N}{4}] are produced closer to the excitation source, and lower-ionization lines such as [\ion{O}{2}] arise farther out and over a more extended region, whereas [\ion{S}{2}] is only produced in the outer edge of the nebula \citep{kewley_understanding_2019,berg_characterizing_2021,harikane_jwst_2025}. Our isobaric photoionization models naturally reproduce this density variation.

Beyond its impact on the measurement itself, $n_{\rm e}$ may be intrinsically correlated with N/O through the nitrogen enrichment mechanism. A positive relation between N/O (from optical nitrogen lines) and $n_{\rm e}$ (from the optical [\ion{S}{2}] doublet) has been reported for the CLASSY sample and for $z=2-6$ star-forming galaxies \citep{arellano-cordova_classy_2025}. This trend suggests that density structure contributes to the scatter in the N/O–O/H relation \citep{arellano-cordova_classy_2025} and may indicate that ISM density influences the nitrogen-enrichment mechanism, which subsequently affects the observed N/O ratio. Our high-$z$ `N-enhanced' sample shows a general increase of N/O with gas pressure, albeit with substantial scatter. A larger `N-enhanced' sample will be needed to test whether this correlation extends to AGNs and persists at $z>6$.

Several studies find that N/O ratios derived from UV nitrogen lines are often higher than those inferred from optical diagnostics \citep[e.g.,][]{rivera-thorsen_sunburst_2024, arellano-cordova_classy_2025}, raising the question of whether N/O in high-z `N-enhanced' galaxies is genuinely elevated. For example, in GS 3073 the optical [\ion{N}{2}]/[\ion{O}{2}] ratio implies $\log(\rm N/O)\approx-1.10$ to $-0.58$, whereas the UV lines yield $\log(\rm N/O)=0.42$ \citep{ji_ga-nifs_2024}. This UV–optical offset has been interpreted as evidence for ISM stratification, with higher N/O ratios in denser regions \citep{pascale_nitrogen-enriched_2023,ji_ga-nifs_2024}. However, for the two galaxies in our sample with optical [\ion{N}{2}] detections (GS 3073 and RXCJ2248-ID), our best-fit models reproduce both the optical and UV nitrogen fluxes using a single N/O value. A larger galaxy sample with both UV and optical nitrogen lines observed is needed to determine whether this UV-optical N/O offset reflects true abundance variations or arises from measurement and methodological differences.
}

\subsection{What Causes Nitrogen Enhancement in High-$z$ Galaxies?}

Since the discoveries of high-redshift nitrogen-enhanced galaxies, several physical mechanisms have been proposed to account for their elevated nitrogen abundance. The three mostly discussed mechanisms are nitrogen enrichment via Wolf-Rayet (WR) stars \citep{watanabe_empress_2024,marques-chaves_extreme_2024}, supermassive stars with masses exceeding $M>1000M_{\odot}$ \citep{charbonnel_n-enhancement_2023, senchyna_gn-z11_2023,marques-chaves_extreme_2024}, and super star clusters (SSCs) \citep{pascale_nitrogen-enriched_2023,topping_metal-poor_2024}.

{\color{black} The high gas pressures ($7.0\leq\log (P/k)\leq9.8$), high ionization parameter ($\log(U)\gtrsim-2.0$), and compact morphology ($R_e \lesssim 120\,$pc) of our `N-enhanced' sample favor the SSC scenario. SSCs are very young (ages $\lesssim10\,$Myr), massive ($M\gtrsim10^4M_{\odot}$), compact ($r\sim \rm pc$), and dense ($\rho\gtrsim10^3M_{\odot}/\rm pc^3$) stellar systems whose intense, localized feedback can strongly reshape their surrounding ISM. A compelling low-redshift analogue of these high-redshift systems is the Sunburst Arc, where strong lensing reveals a massive ($M\gtrsim10^5M_{\odot}$) dense cloud with extremely high gas pressure ($\log(P/k)\approx9.6$) and elevated nitrogen abundance ($\log(\rm N/O)=-0.21$), adjacent to a metal-poor ($Z\sim0.22Z_{\odot}$) SSC of $M\gtrsim10^7M_{\odot}$ and age$\lesssim4\,$Myr \citep{pascale_nitrogen-enriched_2023}. This system provides direct evidence for localized nitrogen enrichment driven by SSC feedback. Follow-up rest-frame optical spectra further reveal WR features in the Sunburst Arc \citep{rivera-thorsen_sunburst_2024}, confirming the presence of WR stars in the N-enhanced system.

SSCs are known to host WN-type WR binaries in the Milky Way \citep{zwart_young_2010}. As birthplaces of WR stars and massive stars ($M\gtrsim60M_{\odot}$), SSCs could produce nitrogen-rich stellar ejecta through the stellar wind from WR stars or the collapse of massive stars to enrich the nitrogen abundance of the surrounding ISM. The underlying pathway is the CNO cycle, which fuses hydrogen into helium using carbon and oxygen as catalysts and builds up nitrogen as a byproduct via the rate-limiting nitrogen-to-oxygen conversion step. When massive stars loss mass through stellar winds after the main sequence (e.g., WR stars), or end their lives through collapse at the end of the main sequence (e.g., supermassive stars with $M\sim1000M_{\odot}$, see \citet{charbonnel_n-enhancement_2023,senchyna_gn-z11_2023}), the nitrogen-rich, oxygen- and carbon-depleted stellar materials are ejected into the surrounding ISM, resulting in elevated N/O and lower C/O ratios in our sample. 

Evidence for WR stars in these ``N-enhanced'' high-redshift galaxies is supported by our photoionization modeling: reproducing the observed UV \ion{He}{2} fluxes in the \ion{H}{2} region models requires including WR star spectra in the FSPS while calculating the stellar ionizing spectra (Section~\ref{sec:3.1}). When the WR component is removed, the predicted \ion{He}{2} lines are $\sim2\,$dex fainter than observed. WR stars are also needed to account for the broad \ion{He}{2} emission. \citet{marques-chaves_extreme_2024} caveats that the nitrogen-rich WN phase in the WR star scenario is short and only lasts for $\sim0.5\,$Myr, before He-burning in the WC phase boosts oxygen and brings down the N/O ratio again. However, progression from WN to the WC phase is strongly metallicity dependent and is suppressed at low metallicity \citep[e.g.][]{eldridge_implications_2006,mokiem_vlt-flames_2006,crowther_physical_2007}. Thus, many WR stars in these metal-poor systems may never reach the WC phase, alleviating the timescale tension and allowing elevated N/O to persist.

Supermassive stars ($M>1000M_{\odot}$) could also form in the densest SSCs and contribute to nitrogen enhancement, which was previously proposed to explain the abundance anomalies in globular clusters and massive star clusters \citep{gieles_concurrent_2018}. These mostly convective supermassive stars undergo H-burning in their hot core and eject nitrogen-enriched, C and O-depleted material to the surrounding ISM before they collapse due to general relativistic instabilities at the end of the main sequence \citep{charbonnel_n-enhancement_2023, senchyna_gn-z11_2023,marques-chaves_extreme_2024,nandal_evolution_2024,nandal_evolution1_2024,nandal_explaining_2024,nandal_fast-rotating_2024}. However, the existence of such supermassive stars remains theoretical, with no direct observational evidence currently available.

Stellar winds or outflows in SSCs could drive radiative shocks when they interact with the surrounding ISM, and shock excitation can mimic several AGN-like UV features. A composite power source of star formation and shocks excitation is therefore plausible in some of our high-$z$ and N-enhanced galaxies, particularly GHZ2 and GN-z9p4, whose spectra are best fit by weak AGN models and lack independent AGN features. Combinations of shock and \ion{H}{2} region models could reproduce the positions of these galaxies on several UV diagnostic diagrams (Zhu et al. 2025c, in prep.). However, pure shock models fail to simultaneously match the observed UV line ratios and equivalent widths in our sample (Zhu et al. 2025c, in prep.), implying that shocks alone are insufficient and that AGN photoionization is required in most cases.

The frequent AGN incidence among `N-enhanced' galaxies also raises the possibility that nuclear stellar activity may seed both AGN growth and nitrogen enrichment. This is suggested by \citet{isobe_jades_2025}, whose study on the stacked spectra of a sample of $z=4-7$ galaxies (20 broad-line AGNs, 18 narrow-line AGNs, and 665 other galaxies) reveals UV \ion{N}{3}] quintuplet lines only in the stacked spectra of the 20 broad-line AGNs. However, our results show that nitrogen-enhanced galaxies can also host Type 2 AGNs (e.g., GN-z9p4 and GS-z9-0) or be purely star-forming (e.g., RXCJ2248-ID). Moreover, at fixed N/O, AGN photoionization yields brighter UV nitrogen lines than \ion{H}{2} photoionization, and current JWST observations have limited depths that bias detections toward AGN \citep{zhu_only_2025}. A larger statistical sample of nitrogen-enhanced galaxies will be essential to robustly test whether nitrogen enrichment is systematically connected to AGN activity.
}

\section{Conclusions}

This paper presents an investigation of the properties of high-$z$ nitrogen-enhanced galaxies discovered in JWST observations. To properly account for the elevated N/O ratio in these galaxies, we develop a set of nitrogen-enhanced AGN and \ion{H}{2} photoionization models. 

We compare the photoionization models with observational data of local star-forming galaxies, $z<4$ Seyfert galaxies, and high-$z$ nitrogen-enhanced galaxies on the UV diagnostic diagrams. We find that our \ion{H}{2} and AGN photoionization models can effectively reproduce the observed UV emission-line ratios in both local and high-$z$ galaxies. We find the following UV diagnostic diagrams most efficient in constraining galaxy properties:

(1) The \ion{N}{4}]/\ion{N}{3}]-\ion{N}{3}]/\ion{O}{3}] and \ion{N}{4}]/\ion{N}{3}]-\ion{N}{4}]/\ion{O}{3}] UV diagnostic diagrams are most effective in identifying galaxies that have elevated N/O ratios.

(2) The \ion{N}{4}]/\ion{N}{3}]-\ion{C}{3}]/\ion{O}{3}] UV diagnostic diagram is most effective in constraining the gas-phase metallicity of the galaxy, if the carbon abundance follows the local carbon scaling relation.

(3) The EW-based UV diagnostic diagrams are most effective in separating strong AGNs ($\log(E_{\text{peak}}/{\rm keV})\gtrsim -1.85$) and star-forming galaxies with moderate to high metallicity ($12+\log(\rm O/H)\geq8.0$). 

However, the ambiguity in distinguishing between soft AGN ($\log(E_{\text{peak}}/{\rm keV})\lesssim -1.85$) and metal-poor ($12+\log(\rm O/H)\leq8.0$) star-forming galaxies remains, because their predicted emission-line ratios overlap significantly in all UV diagnostic diagrams. Most of the high-$z$ nitrogen-enhanced galaxies fall within these ambiguous regions. To further constrain the dominant excitation source of these nitrogen-enhanced galaxies, we include optical observations and perform a comprehensive cross-match between model predictions and observed emission line fluxes and EW measurements to determine the best-fit model for each galaxy. These best-fit models show that:

(1) {\color{black}Of the eight nitrogen-enhanced galaxies, seven (GHZ9, GS3073, GN-z11, GHZ2, CEERS-1019, GN-z9p4, and GS-z9-0) host AGNs. GHZ9 and GS3073 exhibit moderately hard AGN radiation fields, whereas GN-z11, GHZ2, CEERS-1019, GN-z9p4, and GS-z9-0 are consistent with weak AGN models (although partial contribution from star formation can not be excluded). The remaining source, RXCJ2248-ID, is best explained by pure star formation.} 

(2) All these galaxies have moderate to high gas pressure $7.0\leq\log (P/k)\leq9.8$ (corresponding to electron densities in the range of $10^3\leq n_{\rm e} (\rm cm^{-3})\leq10^6$) and high ionization parameter ($\log(U)\gtrsim-2.0$). This suggests that these nitrogen-enhanced galaxies are dense and compact. 

Given the dense, compact properties of high-$z$ nitrogen-enhanced galaxies, we suggest that super star clusters containing WR and massive stars are the most likely contributors to nitrogen enrichment. The prevalence of AGNs in these high-$z$ galaxies also highlights that AGN models should be included in the analysis of high-$z$ galaxies to account for a wider variety of possible excitation sources.

\section{Acknowledgement}

We thank the anonymous referee for thoughtful and useful comments, which have significantly improved this paper. P.Z. would like to thank Charlie Conroy for helpful discussions on the development of this paper. JAAT acknowledges support from the Simons Foundation and \emph{JWST} program 3215. Support for program 3215 was provided by NASA through a grant from the Space Telescope Science Institute, which is operated by the Association of Universities for Research in Astronomy, Inc., under NASA contract NAS 5-03127.

This work is based in part on observations made with the NASA/ESA/CSA James Webb Space Telescope. The data were obtained from the Mikulski Archive for Space Telescopes at the Space Telescope Science Institute, which is operated by the Association of Universities for Research in Astronomy, Inc., under NASA contract NAS 5-03127 for JWST. This work also uses observations made with the NASA/ESA Hubble Space Telescope obtained from the Space Telescope Science Institute, which is operated by the Association of Universities for Research in Astronomy, Inc., under NASA contract NAS 5–26555.

\bibliography{Paper1_UVD_accepted(12:3:25)/Paper4}{}
\bibliographystyle{aasjournal}

\end{document}